%% file: ms.tex
\documentclass[iop]{emulateapj}

\usepackage{graphicx}
\usepackage{hyperref}
\newcommand{\Reyn}{\ensuremath{\mathrm{Re}}}
\newcommand{\Rmag}{\ensuremath{\mathrm{Rm}}}
\newcommand{\Rmagc}{\ensuremath{\mathrm{Rm}_\mathrm{crit}}}
\newcommand{\Prandtl}{\ensuremath{\mathrm{Pm}}}
\newcommand{\Ukep}{\ensuremath{\mathbf{U}^{\mathrm{kep}}}}
\newcommand{\Phikep}{\ensuremath{\phi_{\mathrm{kep}}}}
\newcommand{\alphaSS}{\ensuremath{\alpha_{SS}}}
\begin{document}
\title{Magnetorotational turbulence transports angular momentum in stratified disks with low magnetic Prandtl number but magnetic Reynolds number above a critical value}
\author{Jeffrey S. Oishi\altaffilmark{1}}
\affil{Kavli Institute for Particle Astrophysics and Cosmology,
  Stanford University}
\affil{2575 Sand Hill Road M/S 29, Menlo Park, CA 94025, USA}
\altaffiltext{1}{also at Department of Physics, Stanford University, Stanford,
CA 94305}
\email{jsoishi@stanford.edu}
\and
\author{Mordecai-Mark Mac Low}
\affil{Department of Astrophysics, American Museum of Natural History}
\affil{Central Park West at 79th St, New York, NY 10024, USA}
\email{mordecai@amnh.org}

\begin{abstract}
The magnetorotational instability (MRI) may dominate outward transport
of angular momentum in accretion disks, allowing material to
fall onto the central object. Previous work has established that the
MRI can drive a mean-field dynamo, possibly leading to a
self-sustaining accretion system. Recently, however, simulations of
the scaling of the angular momentum transport parameter $\alphaSS$
with the magnetic Prandtl number $\Prandtl$ have cast doubt on the
ability of the MRI to transport astrophysically relevant amounts of
angular momentum in real disk systems. Here, we use simulations
including explicit physical viscosity and resistivity to show that
when vertical stratification is included, mean field dynamo
action operates, driving the system to a configuration in which the
magnetic field is not fully helical. This relaxes the 
constraints on the generated field provided by magnetic helicity
conservation, allowing the generation of a mean field on timescales
independent of the resistivity. Our models demonstrate the existence
of a critical magnetic Reynolds number $\Rmagc$, below which transport
becomes strongly $\Prandtl$-dependent and chaotic, but above which the
transport is steady and $\Prandtl$-independent. Prior simulations
showing $\Prandtl$-dependence had $\Rmag < \Rmagc$. We conjecture that
this steady regime is possible because the mean field dynamo
is not helicity-limited and thus does not depend on the details of the
helicity ejection process. Scaling to realistic astrophysical
parameters suggests that disks around both protostars and
stellar mass black holes have $\Rmag >> \Rmagc$. Thus, we suggest that
the strong $\Prandtl$ dependence seen in recent simulations does not
occur in real systems.
\end{abstract}

%\pacs{95.30.Qd}

\maketitle
\section{Introduction}
Accretion disks form in a wide variety of astrophysical objects, from
active galactic nuclei to dwarf novae and protostars. In the former
two, disks are responsible for the tremendous luminosities, while in
the latter they are the site of planet formation. Understanding the
structure and evolution of disks is therefore critical to
understanding each of these objects. The successful
\citet{1973A&A....24..337S} model of accretion disks relies on viscous
transport of angular momentum outward through the disk, allowing mass
to spiral inwards. The amount of viscosity necessary to explain
observations requires turbulence, as molecular viscosity is far too
small.  \citet{1973A&A....24..337S} parametrized the turbulent stress
providing this viscosity by $\alphaSS = (\left< \rho u_x u_y\right> -
\left<B_x B_y\right>)/P_0$ where $P_0$ is the midplane pressure of the
disk. The magnetorotational instability (MRI) offers the most viable
mechanism for providing these enhanced levels of angular momentum
transport \citep{1998RvMP...70....1B}. It is known to saturate in a
magnetohydrodynamic turbulent state that transports angular momentum
outward through the disk.

Aside from its central role in accretion disk theory, the MRI also
provides an interesting system in which to study mean field dynamo
theory. Starting from a zero-net flux magnetic field configuration,
the MRI acts as a mean field dynamo, generating strong, fluctuating
fields with order at wavelengths as large as the simulation box
\citep{1995ApJ...446..741B, 1996ApJ...464..690H}.  The system, fed by
the free energy from Keplerian shear, can sustain angular momentum
transport with only a weak (sub-equipartition) field and the presence
of a negative radial gradient in angular velocity. Because the
turbulence is driven by the MRI itself, the Lorentz force is essential
to the operation of the dynamo, ensuring that the system is never in a
kinematic phase \citep{1996ApJ...464..690H}.

In recent years, the MRI has become the target of intense numerical
investigation owing to recent studies showing a decline of the angular
momentum transport rate with increasing resolution
\citep{2007A&A...476.1113F, 2007ApJ...668L..51P} in the simplest 3D
systems that demonstrate MRI turbulence: unstratified, periodic, shearing
boxes. \citet{2007MNRAS.378.1471L} and \cite{2007A&A...476.1123F}
showed that this decline can be described as a rather steep
power-law dependence of transport on the magnetic Prandtl number,
$\Prandtl \equiv \nu/\eta$, the ratio of viscous momentum diffusion to
resistive magnetic diffusion. Given that real astrophysical disks,
especially protoplanetary disks, have extremely low $\Prandtl \sim
10^{-8}$ \citep{2008ApJ...674..408B}, this would imply that the MRI
could not be responsible for angular momentum transport in such systems.

Subsequently, several authors have attempted to explain these angular
momentum scaling results, and in doing so suggest why they may not be
applicable to accretion disks. These include the idea that
unstratified shearing boxes lack a characteristic outer scale
\citep{2009ApJ...696.1021V}; that the initial magnetic field strength
in the \citet{2007A&A...476.1123F} simulations is too weak and they
are thus stable to non-axisymmetric MRI modes that are essential to
sustained transport \citep{2010A&A...513L...1K}; or that an
unstratified shearing box with periodic boundary conditions cannot
sustain large scale dynamo action, and small-scale dynamo action is
known to be $\Prandtl$-dependent, thus rendering the magnetic field
necessary for the MRI susceptible to similar $\Prandtl$-dependence
\citep{2010arXiv1004.2417K}. A pair of recent papers has demonstrated
that even without explicit viscosity and resistivity,
\emph{stratified} MRI simulations \emph{do} converge to a consistent
value of $\alpha_{SS}$ with increasing resolution
\citep{2010ApJ...713...52D, 2010ApJ...708.1716S}. This strongly
suggests that stratification plays a major role in the dynamics and
saturation of the MRI, and demands a thorough investigation of how
it does so, especially in the case of a zero-net flux field, where
dynamo action is inextricably linked to continued accretion.

In the absence of an externally imposed mean field, the shearing box
system can completely destroy the initial magnetic field. For
sustained turbulence to be possible, the MRI must create field by
dynamo action. This field in turn allows the continued excitation of
new MRI modes. Thus, the zero-net flux shearing box is a non-linear
system balancing fluctuating, dynamo-generated fields with MRI
generated turbulence. In order to understand the angular momentum
transport from MRI turbulence and how it scales with the physical
parameters of the system, we need to understand the underlying dynamo
operates.

In this paper, we focus our analysis on the question of dynamo action
in the MRI and attempt to understand the connection between mean field
dynamos and the amount of angular momentum transport. We include
explicit viscosity and resistivity in order to consider the role of
the viscous and magnetic Reynolds numbers $\Reyn$ and $\Rmag$ in
stratified disks. By considering a broader range of $\Prandtl$ than
previous studies, we find evidence for the existence of a critical
magnetic Reynolds number in \emph{stratified} disks. This $\Rmagc$ may
represent a boundary for sustained dynamo activity, which in turn
controls angular momentum transport. We note two dynamo behaviors, one
corresponding to a sustained, organized dynamo, and the other
corresponding to a transient, chaotic dynamo. These behaviors manifest
above and near the critical $\Rmagc$, respectively. We then attempt to
connect the properties of the stratified MRI to other, better studied
dynamo systems at high and low $\Prandtl$. Specifically, we focus on
the conservation of magnetic helicity, and attempt to understand why
the MRI is able to build mean fields on a relatively rapid timescale
compared with other mean field dynamos. We find that the MRI generated
dynamo fields are not significantly helical at any time during their
evolution, and this fact explains why their growth is not limited by
the resistive timescale as might be expected for fully helical fields.

This study is most similar to the work of \citet{2010arXiv1004.2417K},
in that we consider the effects of boundary conditions on the
transport of magnetic helicity and its relation to the generation of
large-scale magnetic fields. However, we consider the effect of
varying $\Prandtl$ on stratified shearing boxes. Furthermore, we
attempt to understand why the \emph{stratified} but \emph{periodic}
simulations of \citet{2010ApJ...713...52D,2010ApJ...708.1716S} show
convergence with resolution. Because of their periodic boundary
conditions, they cannot support the same type of flux-transport dynamo
action that renders angular momentum transport independent of
$\Prandtl$ in the simulations of \citet{2010arXiv1004.2417K}. It is
our conclusion that, despite the presence of a large scale dynamo, the
large scale field is not helicity limited, though the MRI does eject
helicity through open boundaries if they are present. This 
explains why the MRI dynamo can operate in periodic boxes, and
points toward a theoretical understanding of the non-linear shearing
box system. Additionally, the dynamo period does not show a discernible dependence
on $\Prandtl$.  Most importantly, we find that the angular momentum
transport coefficient appears independent of $\Prandtl$ \emph{as long
  as the magnetic Reynolds number $\Rmag$ remains above a critical
  value, $\Rmag_{crit} \sim 3000$ }.

In \S~\ref{s:methods}, we review our methods. We present the scaling
results in \S~\ref{s:results}, followed by a discussion of the
importance of magnetic helicity conservation for these results in
\S~\ref{s:discussion}. We conclude and note several avenues for future
work in ~\S~\ref{s:conclusions}.

\section{Methods}
\label{s:methods}
Using the shearing box formalism
\citep{1995ApJ...440..742H,2008A&A...481...21R}, we study a stratified
patch of a Keplerian accretion disk threaded by an initial magnetic
field of the form $\mathbf{B_0} = B_0 \sin(x) \mathrm{\mathbf{e_z}}$
with maximum midplane value of plasma $\beta = 2 c_s^2/v_A^2 \simeq
89$. We solve the equations of isothermal, compressible
magnetohydrodynamics using the Pencil
Code\footnote{\texttt{http://www.nordita.org/software/pencil-code/}}
\citep{2002CoPhC.147..471B, 2009ApJ...697.1269J}, a spatially
sixth-order, temporally third-order finite difference method. The constraint $\mathbf{\nabla \cdot B} = 0$ is enforced by solving the
evolution equation for the magnetic vector potential,
\begin{equation}
  \label{e:induction}
  \frac{\partial \mathbf{A}}{\partial t} + u_y^0 \frac{\partial
    \mathbf{A}}{\partial y} =
  \mathbf{u \times B} + \frac{3}{2} \Omega A_y \mathbf{\hat{e}_x+ \eta \nabla^2 A + \eta_3 \nabla^6 A} + \mathbf{\nabla} \Phikep,
\end{equation}
where $\eta_3 \mathbf{\nabla^6 A}$ is a hyperdiffusion operator to
dissipate excess energy at the grid scale, and the second terms on
each side are from the shearing box formalism. We use similar
hyperdiffusion terms on all dynamical equations
\citep[see][]{2005ApJ...634.1353J,2007ApJ...670..805O}. The value of
$\eta_3$ given in Table~\ref{t:runs} is chosen so that the
hyperdiffusive Reynolds numbers are roughly unity at the Nyquist
scale, where $\Reyn_{Ny} = u_{\mathrm{rms}} / \eta_3 k_{ny}^{5}$. the
magnitude of the hyperdiffusion operators is chosen for numerical
stability and speed, and has no significant effects on our
results. The hyperdiffusivity scales inversely with the grid size dx,
so the convergence seen in our high resolution runs suggests that
$\Rmagc$ is unlikely to be affected by the value of the
coefficient. 

\begin{figure}
  \includegraphics[width=0.95\columnwidth]{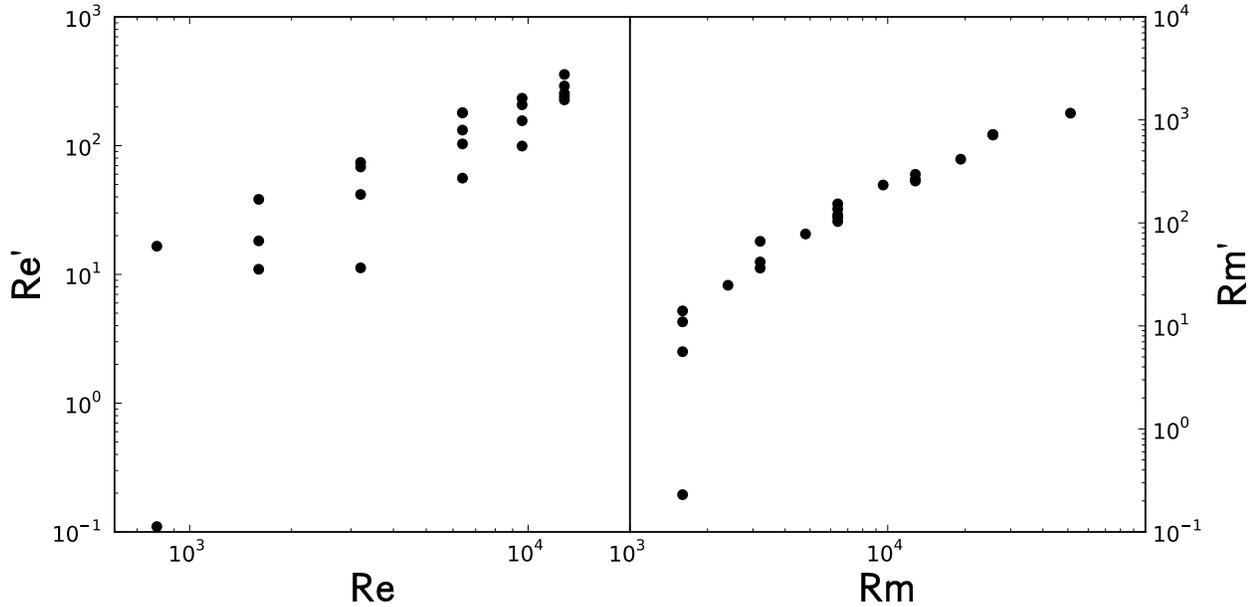}
  \caption{(left) Reynolds number parameter $\Reyn$ versus the ratio of advection to viscous dissipation in the simulations. (right) Magnetic Reynolds number parameter $\Rmag$ versus the ratio of advection to resistive dissipation in the simulations. \label{f:re_rm_corr}}
\end{figure}

In order to study the effects of $\Prandtl$ on the saturated MRI
turbulence, we vary the viscosity $\nu$ and resistivity $\eta$. We
define $\Reyn \equiv S H^2/\nu$ and $\Rmag \equiv S H^2/\eta$, where
$S = \Omega d \log \Omega/ d \log r = q \Omega = -3/2 \Omega$ is the
(Keplerian) shear rate in the box in units of rotation $\Omega$ and
$H^2 = c_s^2 / \Omega^2$ is the scale height of the disk. Note that
our definition of $S$ is opposite in sign with respect to
\citet{2007MNRAS.378.1471L}; there is no consistent definition. We set
$c_s = H = \Omega = \mu_0 = 1$, choosing our units to minimize the
number of values we need to remember. In these units, $\Reyn =
1.5/\nu$, $\Rmag = 1.5/ \eta$, and $B_0 = 0.15$. The advantage of
defining $\Reyn$ and $\Rmag$ this way is that they can be set \emph{a
  priori}, though they do not measure the relative effects of
advection and dissipation in the saturated state of the MRI. Because
our $\Reyn$ and $\Rmag$ are \emph{a priori} parameters, we have
checked that they correlate with the ratio of turbulent advection to
dissipation, $\Reyn' = u_{rms}/k_1 \nu$ and $\Rmag' = u_{rms}/k_1
\eta$, respectively. Here, $k_1 = 2\pi/L_z$ is the smallest integer
wavenumber in the box. Note that $k = k_1/2$ is consistent with the
vertical field boundary conditions, and large-scale fields of this
size can also fit within the box. This data is presented in
Table~\ref{t:runs}. Indeed, in Figure~\ref{f:re_rm_corr} $\Rmag$ is
particularly well correlated with $\Rmag'$.

Table~\ref{t:runs} summarizes the parameters of our models. We report
resolution in terms of zones per scale height, with a standard
resolution of $64\ \mathrm{zones/H}$, though we also ran three
simulations with $128\ \mathrm{zones/H}$ to confirm the convergence of
our results. All of our runs use cubic zones, $dx = dy = dz$ and were
run for $100 t_{orb}$. Our standard box size is $1 \mathrm{H} \times 4
\mathrm{H} \times 4 \mathrm{H}$. 

Our simulations are periodic in $y$ (azimuthal), shearing periodic in
$x$ (radial), and one of three different choices for $z$ (axial):
periodic, perfect conductor, or vertical field (hereafter VF). Among
these choices, the first does not allow a flux of magnetic helicity
out of the simulation domain, while the others do. This has a
significant effect on the resulting dynamo action and turbulence,
though not nearly as dramatic as in the unstratified results of
\citet{2010arXiv1004.2417K}. Vertical field boundary conditions ensure
$B_x = B_y = 0$ on the upper and lower boundaries. They have been used
before in a number of accretion disk
\citep[e.g.][]{1995ApJ...446..741B, 2001A&A...378..668Z,
  2010arXiv1004.2417K, 2010MNRAS.405...41G} and magnetoconvection
\citep{1988ApJ...327..920H} simulations as a simplified version of the
vacuum boundary conditions appropriate to the surface of a disk or
star. The total magnetic flux in this case is not conserved, but is
free to grow or decay in the $B_x$ and $B_y$ components, thus allowing
a mean field dynamo action within the domain in a computationally
convenient manner, though they provide a somewhat artificial constraint
on the field at the boundaries. They are known to produce spurious
current density near the boundaries \citep{2001A&A...378..668Z}, but
this effect does not significantly affect bulk properties of the flow,
as we show below. The vertical field conditions are implemented in our
vector potential code by setting $\partial_z A_x = \partial_z A_y =
A_z = 0$ at the $z$ boundaries.

\section{Results}
\label{s:results}
Here, we present the main results from the suite of simulations
described in Table~\ref{t:runs}. 

\subsection{Scaling with $\Prandtl$ and $\Rmag$}
\begin{figure}[h]
  \includegraphics[width=0.7\columnwidth]{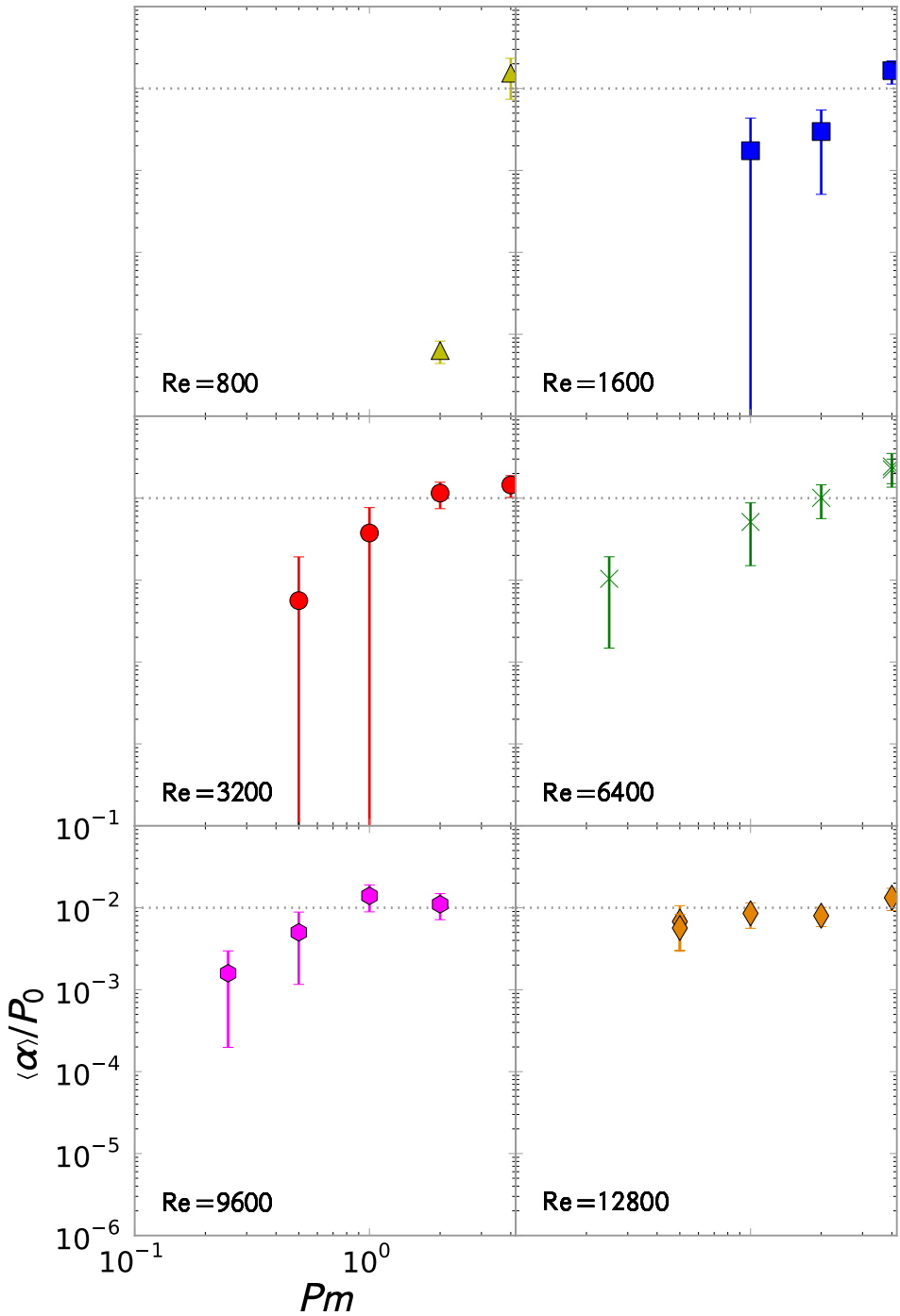}
  \caption{Volume and time averaged value of the angular momentum
    transport parameter $\alphaSS$ as a function of magnetic Prandtl
    number $\Prandtl$. Each subfigure is plotted on the same axes, but
    shows models with different Reynolds number $\Reyn$. Angular
    momentum transport increases with $\Prandtl$ up to a threshold
    value of $\Prandtl$ beyond which it remains constant. This
    threshold occurs at lower $\Prandtl$ for increasing $\Reyn$,
    suggesting a constant threshold at some critical magnetic Reynolds
    number $\Rmagc$. The superposed points in the $\Reyn = 6400$,
    $\Reyn=9600$ (not visible), and $\Reyn = 12800$ series represent resolution
    studies with one point representing a model with 128 zones/$H$,
    double the standard resolution.}
  \label{f:alpha_vs_pm}
\end{figure}

Figure~\ref{f:alpha_vs_pm} shows that the dependence of
$\alphaSS$ on $\Prandtl$ is not well described by a single power-law
at a given $\Reyn$, as had been previously claimed for unstratified
shearing boxes by \citet{2007MNRAS.378.1471L} and \citet{2007A&A...476.1123F}. The
points are the mean $\alpha_{SS}$ for all times $t > 20 t_{orb}$, and
the error bars represent the standard deviation over that range. These
runs all use the VF boundary conditions. The figure shows evidence of
a cut-off that moves to higher $\Prandtl$ as $\Reyn$ decreases. This
is indicative of a critical $\Rmag$, most visible for $\Reyn = 3200$
(center left) and $\Reyn = 9600$ (lower left). Above $\Rmagc$, it
appears that $\alphaSS$ is consistent with a constant value. These
data show that the behavior of the stratified, zero net flux MRI
system has two distinct regimes, controlled by $\Rmagc$. When $\Rmag <
\Rmagc$, the transport is strongly dependent on $\Prandtl$; when
$\Rmag > \Rmagc$, the transport is independent of $\Prandtl$. Given
that astrophysical disks have typical $\Reyn \gtrsim 10^{16}$, this
result means that the MRI is capable of robust angular momentum
transport even at low $\Prandtl$, so long as $\Rmag > \Rmagc$. We
discuss these estimates in more detail in \S~\ref{s:conclusions}
and suggest that this criterion is easily met in most disks.

\begin{figure}
  \includegraphics[width=0.95\columnwidth]{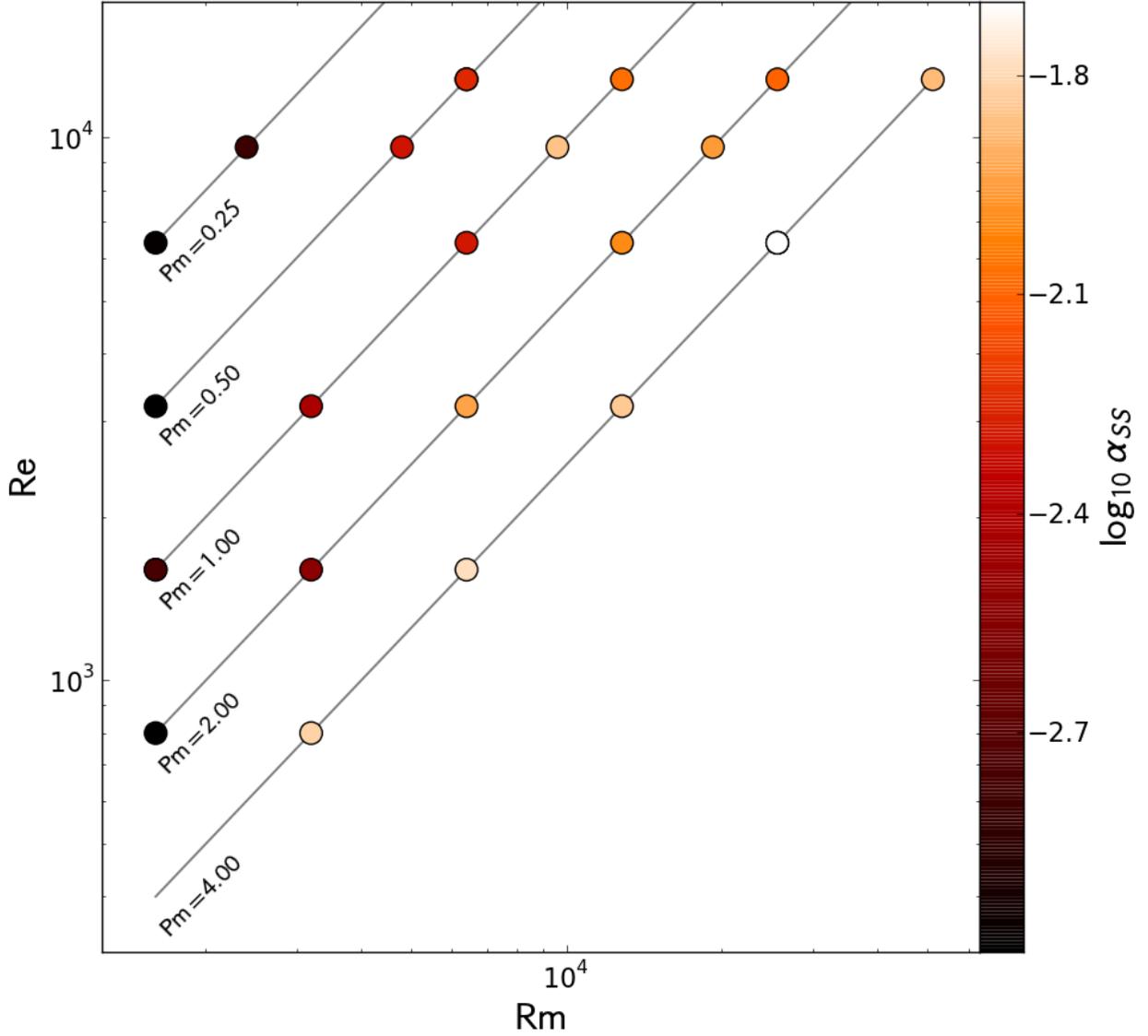}
  \caption{The log of $\alpha_{SS}$ as a function of both viscous and
    magnetic Reynolds numbers $\Rmag$ and $\Reyn$. Darker shades
    represent weaker angular momentum transport. Below a critical
    magnetic Reynolds number $\Rmagc$ slightly above $10^3$, transport
    falls off significantly.}
  \label{f:re_rm_alpha_grid}
  
\end{figure}

We recast these results in the $(\Rmag, \Reyn)$ plane in
Figure~\ref{f:re_rm_alpha_grid}. This figure makes clear the fact that
along with a reduced but still-present $\Prandtl$-dependence, there
is a fairly clear critical $\Rmag$: transport is
significantly reduced left of the vertical line near $\Rmag \sim
3000$. As the simulations approach $\Rmagc$, the $\alphaSS$
variance increases significantly. 

\subsection{Two Dynamo Behaviors}
Figure~\ref{f:bym_bxm_vs_t} shows that the MRI appears to operate in
two distinct dynamo states, one in which regular $\left< B_y \right>$
cycles appear with a characteristic period of $\tau_B \sim 10
t_{orb}$, and one with irregular variations with timescales on the
order of $\sim 50 t_{orb}$. The key control parameter appears to be
$\Rmag$, as the second, third, and fourth panels from the top show the
regular behavior despite being at different $\Prandtl$. There is no
discernible trend in cycle period with $\Prandtl$, though the top
panel, with $\Rmag = 12800, \Prandtl=2$ seems to show an intermediate
behavior. However, all runs with $\Rmag \lesssim 3200$ conclusively
show the irregular behavior. There appears to be a secondary $\Prandtl$
effect as well, since the two $\Rmag = 3200$ models show different behavior
depending on $\Prandtl$ (third and fifth panels from the
top).  

These two drastically different behaviors complicate efforts to
ascertain scaling properties of the MRI as a function of dimensionless
parameters. The range of available $\Prandtl$ is limited by numerical
resolution both from above and below: large values of $\Prandtl$ imply
a viscous scale much larger than the resistive one, while small
$\Prandtl$ implies the opposite. Since both length scales must fit on
(and be resolved by) the grid, and indeed neither can be so large as
to stabilize the largest linear MRI modes that fit in the simulation
box, our parameter range is necessarily limited. Furthermore, since
low $\Rmag$ runs transition to a very different behavior with much
larger cycle period and very different turbulent properties in a
discontinuous fashion, our range of available $\Prandtl$ space for the
regular dynamo mode is further limited.

Thus, in what follows, we briefly describe the irregular dynamo before
focusing on connecting the details of the MRI turbulence in the
regular region of parameter space to better-established results on
small and large-scale dynamo action. The MRI is known to produce both
small and large scale dynamo action, the latter typically requiring
stratification or open boundary conditions (though
\citet{2008A&A...488..451L} demonstrate a large scale dynamo with
neither). The small-scale dynamo for driven, isotropic, homogeneous,
incompressible turbulence has $\Rmagc$ that \emph{increases} with
\emph{decreasing} $\Prandtl$ \citep{2005ApJ...625L.115S}, and thus it
behooves us to understand how MRI turbulence fits into this picture.

\subsection{Irregular Regime}

When $\Rmag$ drops below $\Rmagc$, the mean field dynamo switches to a
irregular regime, showing quasi-periodic magnetic cycles, with longer
cycle lengths (one hesitates to call them periods) $\sim 40 t_{orb}$
(see the lower two panels of Figure~\ref{f:bym_bxm_vs_t}). This is
consistent with Figure 15 of \citet{2010ApJ...713...52D}, who also
showed long cycle lengths in highly resistive simulations. In this
regime, the turbulence often appears only in one half-plane, either $z
> 0$ or $z < 0$, sometimes staying that way for nearly the duration of
the simulation. This kind of behavior at first appears unphysical, as
the stratified shearing box system, while having odd parity about the
midplane, should not necessarily damp perturbations of one helicity
more than the other. Indeed, in all simulations, the kinetic helicity
shows erratic fluctuations of sign with respect to the midplane.

However, this is explainable in a simple $\alpha - \Omega$
treatment. The only prerequisite for this explanation is that the two
half-planes be dynamically decoupled from one another. For the case of
VF boundary conditions, $B_y(z = \pm L_z) \mapsto 0$. The largest
wavenumber modes compatible with this boundary condition are $B_y
\propto \sin(2 \pi/L_z z)$ and $B_y \propto \sin(\pi/L_z z)$,
corresponding to $k= 1$ and $1/2$, respectively. The $\alpha - \Omega$ dynamo operates when the mean induction equation takes the form 
\begin{equation}
  \partial_t \langle \mathbf{B} \rangle = \alpha \langle \mathbf{B} \rangle - q \Omega \langle B_x \rangle \mathbf{\hat{y}} + \eta \nabla^2 \langle \mathbf{B} \rangle.
\end{equation}
If during a cycle
period, the MRI turbulence shuts off, the $\alpha$ effect that it produces will cease,
leaving a mean induction equation that looks like
\begin{equation}
  \partial_t \left<B_x\right> = \eta \nabla^2 \left<B_x\right>
\end{equation}
for the $x$ component and 
\begin{equation}
  \partial_t \left< B_y \right> = -q \Omega \left< B_x \right> +\eta
  \nabla^2 \left< B_y \right>
\end{equation}
for the $y$ component. Because the decay time for modes is roughly
$t_{decay} \simeq 1/k^2 \eta$, small scale structure will undergo
selective decay, leaving only the largest scale modes, the decay time
of which is $\sim 108 t_{orb}$ for $\Rmag = 1600$. At this stage,
$\left< B_y \right>$ has about this much time to grow linearly via
stretching of any residual $\left< B_x \right>$ to amplitudes at which
non-axisymmetric MRI can reestablish fluid turbulence and hence an
$\alpha$ effect. If the two half-planes are not strongly coupled, it
is possible that the turbulence will die out in one half of the box
first, leading to an $\alpha$ effect only in the upper or lower
midplane. This scenario is essentially the same as the one demonstrated in the
midplane by \citet{2011ApJ...730...94S}.

\subsection{Dynamo coefficients}
Understanding the origin of the scaling of angular momentum transport
with magnetic Prandtl number requires understanding the underlying
field generation mechanism. \citet{2010MNRAS.405...41G} has
demonstrated that the field patterns present in the MRI can be
explained in terms of a mean-field dynamo model that includes
dynamical $\alpha$-quenching, a mechanism that modulates $\alpha$ by
enforcing magnetic helicity conservation as the mean field grows (see
\S~\ref{s:helicity} for more details). Here, we decompose fields from
the simulations into their mean and fluctuating components, $\mathbf{B
  = \bar{B} + b}$, where $\mathbf{\bar{B}}$ is a horizontal ($x-y$)
average and $\mathbf{b}$ is the fluctuating field. We denote full box
averages as $\left<\mathbf{B}\right>$. In the standard $\alpha-\Omega$
dynamo mechanism, the $\alpha$ effect of isotropic, helical turbulence
generates poloidal field from toroidal fields, which are in turn
sheared out by differential rotation $\Omega$ and regenerate toroidal
field, thus leading to exponential amplification. Based on the early
work of \citet{1995ApJ...446..741B}, the traditional $\alpha-\Omega$
scenario can explain the observed periodic dynamo generation and
propagation of $\bar{B_y}$ away from the disk midplane if the $\alpha$
term has the opposite sign of that expected from rotating, stratified
turbulence (the strong Keplerian shear has no problem stretching
poloidal field to generate toroidal field; the issue at hand is
generating the former). However, that expected sign was derived from
an analysis that only assumed the presence of a kinetic $\alpha_{K}$
effect from the helicity of the fluid turbulence. Once the magnetic
$\alpha_{M}$ from the Lorentz force is also included
\citep{1976JFM....77..321P} (see also
  \S~\ref{s:helicity}), the total $\alpha =
\alpha_{K} + \alpha_{M}$ is dominated by $\alpha_{M} = 1/3 \tau
\mathbf{\left< \nabla \times v_a \cdot v_a \right>}$, where $\tau$ is
a typical turbulent correlation time and $\mathbf{v_a = b/\sqrt{4 \pi
    \rho}}$. The total $\alpha$ has the required sign
\citep{2010MNRAS.405...41G}. While we similarly find that $\alpha_{M}
\simeq 10 \alpha_{K}$, our results for the $z$ profile of $\alpha_{M}$
(and thus the total $\alpha$) appear to contradict those of
\citet{2010MNRAS.405...41G}. Figure~\ref{f:alpham_z_sat} shows the
$\alpha_M$ profile for three simulations with $\Prandtl = 1, 4$ and
$\Reyn = 3200, 6400, 12800$. There is no monotonic trend with
$\Prandtl$, and there are some differences in shape, but the overall
profile is negative in the upper plane ($z > 0$) and positive in the
lower, opposite to that found by \citet{2010MNRAS.405...41G}. However,
it is worth noting that both the $\Reyn = 12800$ and the $\Reyn =
3200$ runs show a reversal of the $\alpha_M(z)$ profile near the
midplane, just as found by \citet{2010MNRAS.405...41G}, though with
the opposite sign.

The only significant differences between his study and ours are the
vertical boundary conditions on the fluid, which are outflow in his
case, and the fact that his domain covers $6 H$, while our domains
typically cover $4 H$. We have run one simulation, with $\Reyn = 6400$
and $\Prandtl = 4$, with a vertical domain of $6
H$. Figure~\ref{f:alpham_z_4h6h} shows $\alpha_M(z)$ for both the
standard and extended domain for this model. The figure suggests a
possible explanation for the discrepancy between our models and
Gressel's. When we enlarge our domain, we see regions with $|z|
\gtrsim 2$ showing a positive $\alpha_M$ effect, while the regions
$|z| \lesssim 2$ are roughly the same between the $4 H$ and $6 H$
models, outside of narrow boundary layers in both cases. Thus, our
models just appear to show a stronger reversal of the sign of
$\alpha_M$ near the midplane than does
\citet{2010MNRAS.405...41G}. Given that buoyancy and Parker
instabilities become more prominent with height, it seems that the
most likely explanation for the discrepancy between our results and
Gressel's is a combination of a reduced domain for our results and the
outflow fluid boundary conditions in Gressel's work. The outflow
boundary conditions allow buoyant fluid to escape out of the top of
the box, while our closed, stress free lids force that flow to
recirculate. Thus, we do not consider out apparently discrepant
results to actually be evidence against the standard $\alpha - \Omega$
mechanism acting to drive the observed dynamo phenomenology.

\section{Discussion}
\label{s:discussion}
\subsection{Comparison with previous work}
In order to place our results in the context of other recent studies,
we begin by comparing our results to the unstratified results
presented by \citet{2010arXiv1004.2417K}. These authors performed two
series of unstratified shearing-box simulations of the MRI, both
initialized with zero-net magnetic flux in the $z$-direction. One set
of simulations had periodic boundary conditions on the magnetic field
on the $z$ boundary, the other had VF boundary
conditions. \citet{2010arXiv1004.2417K} show that in the latter case,
the angular momentum transport coefficient $\alphaSS$ is
\emph{independent} of $\Prandtl$, while with periodic boundary
conditions, $\alphaSS \propto \Prandtl^{2}$. We find that even with
periodic boundary conditions, a stratified disk with high enough
$\Rmag$ has $\alphaSS$ basically independent of $\Prandtl$.

We are not the first to suggest a critical $\Rmag$ for MRI
turbulence. In the unstratified case, \citet{2000ApJ...530..464F} and
\citet{2002ApJ...577..534S} both noted the existence of a sharp cutoff
in $\alphaSS$ when the resistivity exceeded a certain value. The
latter study used the Elsasser number, $\Lambda = v_{A0}^2/\eta
\Omega$ (though they refer to this as $\Rmag$). However, aside from
using $\Lambda$, their results also measure the $\alphaSS$ cutoff as a
function of the \emph{initial} magnetic field strength, for both net
flux and zero-net flux initial fields. In our case, we do not make
reference to the initial field strength; instead, by using the
instantaneous value of $\Rmag$ as our parameter, we measure the
relative action of shear against dissipation. This is the appropriate
measure for a non-linear, self-sustained system such as the
stratified, zero-net flux MRI that we study here, because any
dynamo-generated energy must ultimately come from the free energy in
the shear flow.

More recently, a controversy regarding the role of channel modes in
saturating the unstratified MRI when a net $z$ field is present has
led to similar results regarding the $\Rmagc$ for MRI-driven
transport. \citet{2010ApJ...716.1012P} demonstrated the existence of a
critical $\Lambda$ below which the angular momentum transport by the
MRI dropped off dramatically. Numerical simulations of the same
problem by \citet{2010A&A...516A..51L}, though seemingly disproving
the \citet{2010ApJ...716.1012P} saturation mechanism, also suggests a
trend similar to the one we report, with two regimes separated by
$\Prandtl \sim 1$. With $\Prandtl < 1 $, the transport scaling is
dominated by $\Rmag$, while above it the scaling is mostly with
$\Prandtl$. Their primary aim was to elucidate the role of
axisymmetric channel modes in the saturation of MRI turbulence in the
presence of a net $z$ field. In our study, there are no channel modes,
as there is no net $z$ field, and it is not too surprising that our
results are slightly different. Indeed, the physics our control
parameter is attempting to represent is not the growth rate and wave
number of a (quasi-)linear channel mode, but instead a turbulent
dynamo that tends to create a net \emph{toroidal} field ($y$-field in
our simulations), on which any MRI growth is non-axisymmetric and thus
transient though with extremely high growth factors \citep{1992ApJ...400..610B}.

\subsection{Dynamo Action and Current Helicity}
\label{s:helicity}
We interpret our results in light of the ability of the system to
build up a large scale magnetic flux. Our results suggest that, with
stratification, the MRI can act as a mean-field dynamo with any
boundary conditions, unlike the unstratified case. In order to clarify
terms, we will refer to a mean-field dynamo as any system capable of building
magnetic fields at the lowest possible wavenumber in the box, either
$k = 1$ in the periodic case or $k = 0$ (a true mean field) in the case including a VF
boundary condition.

In the modern picture of mean-field dynamo theory, ensuring the conservation of
magnetic helicity,
\begin{equation}
  H \equiv \left< \mathrm{A \cdot B} \right>,
\end{equation}
provides an important constraint on the evolution of the mean
field. Magnetic helicity conservation provides both a compelling
physical mechanism for the existence of mean field dynamos as well as
quantitative predictions for saturated field strengths and the
timescales necessary to reach those strengths. We briefly sketch out
some background related to magnetic helicity conservation and
then apply it to our MRI simulations.

The evolution equation for $H$ can be written
\begin{equation}
  \label{e:helicity}
  \frac{dH}{dt} = -2 \eta \left< \mathbf{J \cdot B} \right> - 2 \oint
  \left( \mathbf{ A \times E} + \phi \mathbf{B} \right) \cdot
  \mathbf{\hat{n}} dS,
\end{equation}
where Ohm's law gives $\mathbf{E = -U \times B} + \eta \mathbf{J}$ and
$\phi$ is an arbitrary gauge term in the definition of the vector
potential \citep{2005PhR...417....1B}. In our simulations we use the
Kepler Gauge, $\phi = \mathrm{\Ukep \cdot A}$; see \S~\ref{s:gauge}
for more details as well as \citet{1995ApJ...446..741B}.  The first term is due
to resistivity acting on the current helicity $C \equiv \left<
\mathrm{J \cdot B} \right>$, while the second is a boundary flux term
that can entirely determine the behavior of the MRI in unstratified
simulations \citep{2010arXiv1004.2417K}. In the stratified case, the
vertical boundary conditions are considerably less important. This is
evident in the periodic, stratified simulations of
\citet{2010ApJ...713...52D}, who find sustained turbulence for lower
$\Prandtl$ than similar unstratified boxes. We note that our gauge
choice eliminates all possible horizontal magnetic helicity fluxes
\citep[see][for a detailed explanation]{2011ApJ...727...11H}, leaving
only the possibility of vertical fluxes, which could be driven by a
turbulent diffusivity or shear, as in the case of
\citet{2001ApJ...550..752V}.

By using an eddy-damped, quasi-normal, Markovian closure scheme,
\citet{1976JFM....77..321P} demonstrated that current helicity drives
a magnetic $\alpha_M$ effect, akin to the fluid $\alpha$ effect of
classical mean field dynamo theory. This $\alpha_M$ in turn leads to
an inverse cascade of magnetic energy, and thus the build up of large
scale magnetic field. That inverse cascade can be seen simply as a
result of the conservation of magnetic helicity. We outline this
process following \citet{2005PhR...417....1B}. Consider a fully
helical field, which has its scale-dependent relative helicity,
\begin{equation}
  \mathcal{H}(k) = \frac{k H(k)}{2 M(k)} = 1.
\end{equation}
$H(k)$ and $M(k)$ are the Fourier transforms of magnetic helicity and
magnetic energy, respectively. $\mathcal{H} = 1$ implies $k H(k) = 2
M(k)$. The non-linear terms allow coupling only among modes
$k$ and $p$ satisfying $p + q = k$, where $q$ is an arbitrary mediating wavenumber in the three-wave interaction, helicity is conserved for each
mode, so $p H(p) = 2 M(p)$ and $q H(q) = 2 M(q)$. Since energy is
conserved in the interaction, $M(k) = M(p) + M(q)$, and thus $p H(p) +
q H(q) = k H(k)$.  Because helicity must also be conserved, $H(k) =
H(p) + H(q)$, and
\begin{equation}
  k = \frac{p H(p) + q H(q)}{H(p) + H(q)}.
  \label{e:invcas}
\end{equation}
If $k$ is the final wavenumber, one of $p$ or $q$ is the starting
wavenumber, and Equation~\ref{e:invcas} demands that $k$ is less than
or equal to the maximum of $p$ or $q$. Since $k$ is
smaller than or equal to the parent wavenumber, this corresponds to an
inverse cascade of magnetic energy. \citet{1976JFM....77..321P}
explicitly relate this to the current helicity, and construct a
magnetic $\alpha_M$ term that back-reacts on the flow as this cascade
proceeds. 

This formalism gives a heuristic explanation for the existence of a
mean field dynamo driven by small-scale, helical flows, and
incorporating the magnetic helicity constraint into mean field dynamo
models allows them to correctly predict the saturation behavior of
$\alpha^2$ dynamos \citep{2001ApJ...550..824B}. Here, we want to
understand the MRI dynamo action in our simulations in light of the
constraints imposed by helicity conservation. We explore both boundary
terms and scale transfer of $H$.

We begin by testing the Pencil Code's numerical conservation of
magnetic helicity. For a system with periodic boundary conditions, the
second term in Equation~\ref{e:helicity} is zero, and the only
contribution to changes in helicity can come from resistively limited
current helicity fluctuations. Therefore, if our simulations
numerically conserve helicity, a run with periodic boundary conditions
should show helicity variations only on a resistive timescale. In
order to establish that helicity conservation is robust in our
simulations, we ran a simulation with $\Reyn = 3200$ and $\Prandtl =
2$, with periodic boundary conditions, and tracked the time evolution
of magnetic and current helicities (upper panel of
Figure~\ref{f:abm}). We used a simple forward Euler scheme to
integrate $dH/dt = -2\eta C$ with volume average data from the
simulations. The $H(t)$ that results from this integration is shown in
Figure~\ref{f:abm} as the green triangles, while the $H$ calculated
directly during the run is given by the blue solid line. The agreement
is quite good, considering the crudeness of the integration method and
the sparseness of the data (it is only sampled at intervals of 100
timesteps). As a result, we are reasonably confident that our
simulations conserve magnetic helicity.

Moving to the case with VF boundary conditions, the lower panel of
Figure~\ref{f:abm} again shows the current helicity and the now-gauge
dependent quantity $H = \left< A \cdot B\right>$. Because of the
presence of the gauge in the second term of Equation~\ref{e:helicity},
this quantity is not physically meaningful itself. However, it is
well-defined and provides a clue as to the behavior of the
system. Once again, we integrate $-2\eta C$ using $C$ in the same
way. We expect that if magnetic helicity ejection occurs as a result
of MRI turbulence, the value of $H$ resulting from this integration will
\emph{not} track the actual value from the simulation, and indeed the
lower panel of Figure~\ref{f:abm} shows that it does not. Because of
our gauge choice, we know that the flux of magnetic helicity flux
density must be vertical, and this figure confirms that significant
amounts of helicity escape. Furthermore, $H$ fluctuates
considerably more frequently in the VF case than in the run with
periodic boundary conditions, showing that the timescale for variation
of the global magnetic helicity decreases considerably when a
helicity flux is allowed.

\begin{figure*}[ht]
  \begin{center}
  \includegraphics[width=0.7\textwidth]{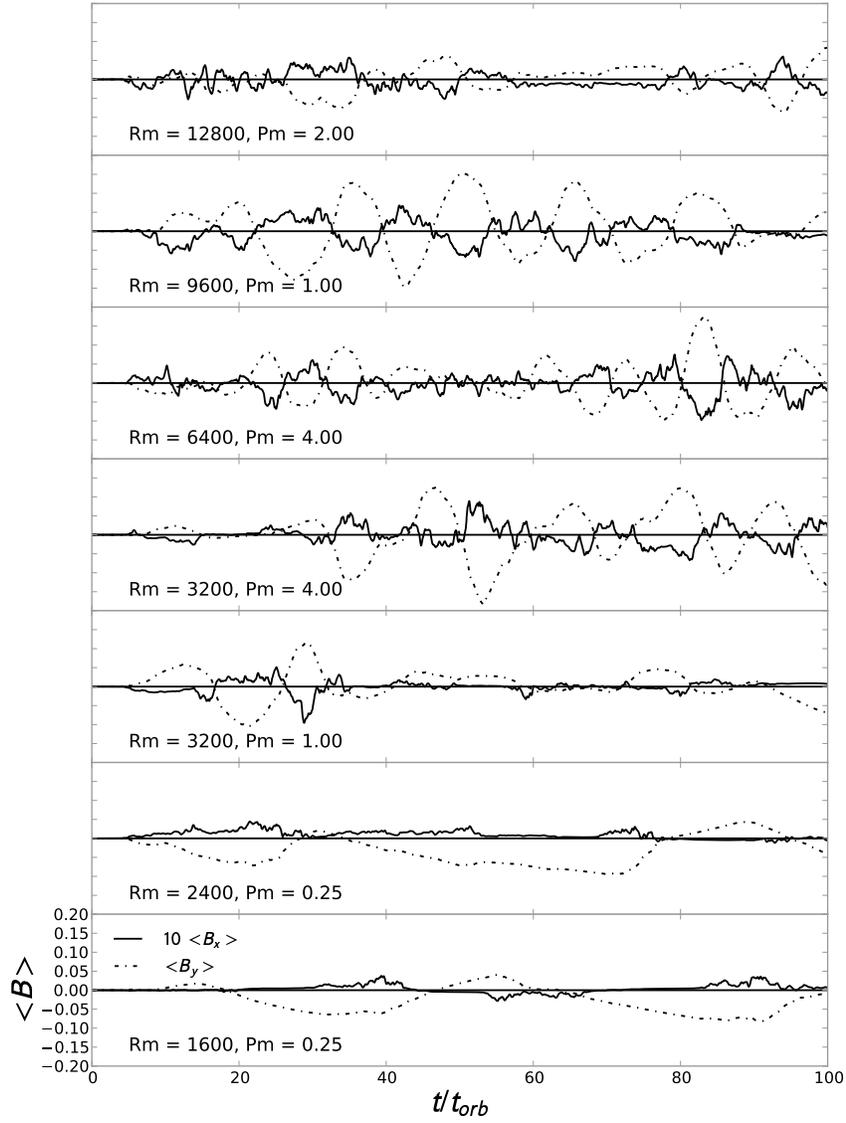}
  \end{center}
  \caption{Mean fields, $\left< B_x \right>$ and $\left< B_y \right>$,
    as a function of time for seven simulations with $\left< B_x
    \right>$ multiplied by a factor of $10$ for clarity. In all cases,
    the $x$ and $y$ components are out of phase, but there are two
    distinct dynamo modes, exemplified by the $\Rmag = 2400, \Prandtl
    = 0.25$ and $\Rmag = 3200, \Prandtl = 4$ cases. All vertical axes for $\left< B
    \right>$ are on the same scale given in the bottom panel.}
  \label{f:bym_bxm_vs_t}
  
\end{figure*}

\begin{figure}
  \includegraphics[width=0.75\columnwidth]{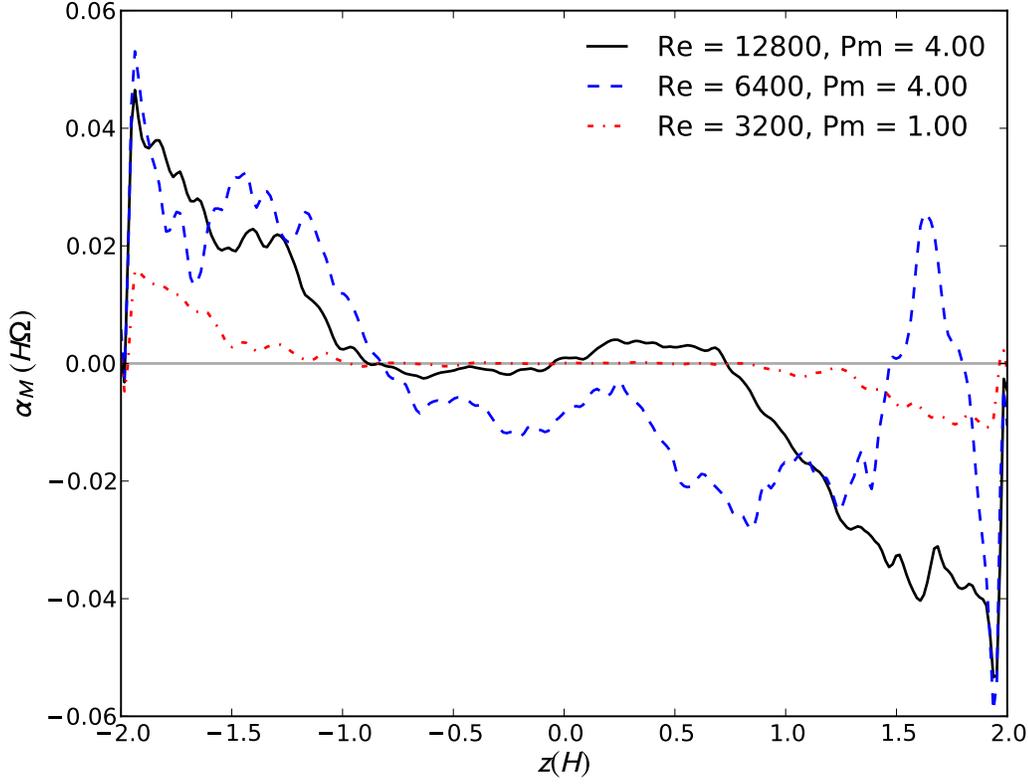}
  \caption{Radial and azimuthially averaged
    $\left<\alpha_M\right>_{xy}(z)$, for several runs averaged over $t
    > 20 t_{orb}$. Note the profile is opposite in sign to that presented in
    \citet{2010MNRAS.405...41G}.}
  \label{f:alpham_z_sat}
\end{figure}

\begin{figure}
  \begin{center}
    \includegraphics[width=0.75\columnwidth]{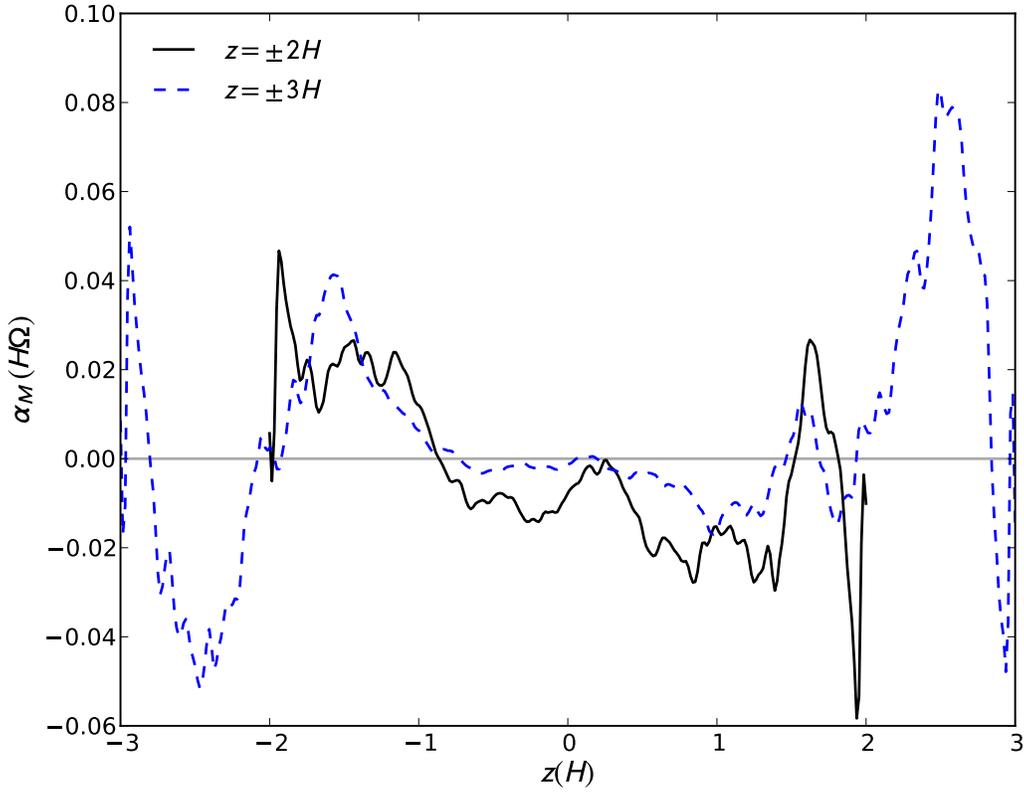}
  \end{center}
  \caption{Radial and azimuthially averaged
    $\left<\alpha_M\right>_{xy}(z)$, for two runs with $\Reyn = 6400$
    and $\Prandtl = 4$ averaged over $t > 20 t_{orb}$. One model has
    the standard vertical extent, $|z| < 2 H$ (solid line), while the other
    has an extended vertical extent, $|z| < 3H$ (dashed line). The sign
    reversal seen above $|z| \sim 2$ may explain the apparent
    discrepancy between our results and those of
    \citet{2010MNRAS.405...41G}.}
  \label{f:alpham_z_4h6h}
\end{figure}

\begin{figure}
\includegraphics[width=0.75\columnwidth]{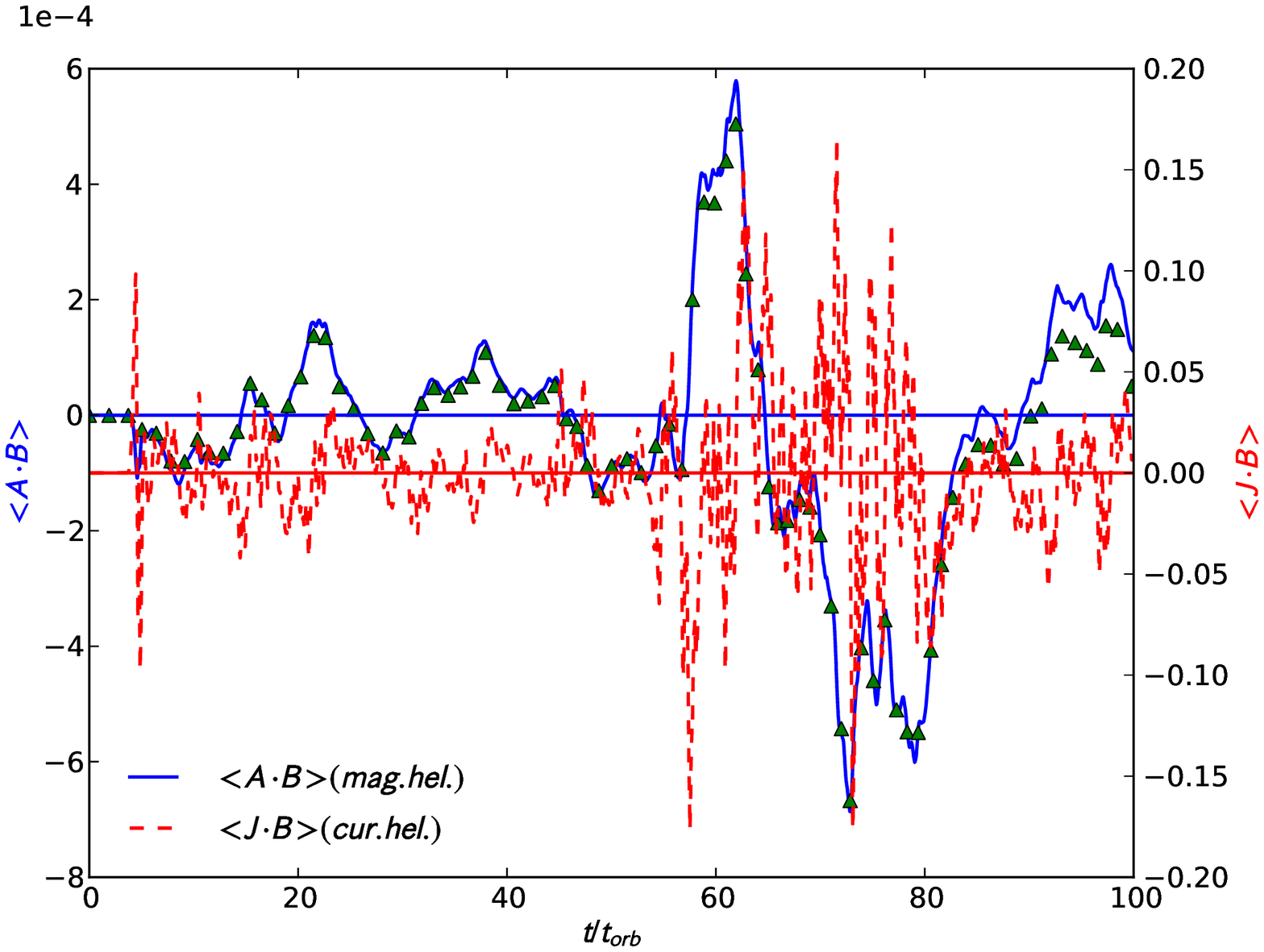}
\includegraphics[width=0.75\columnwidth]{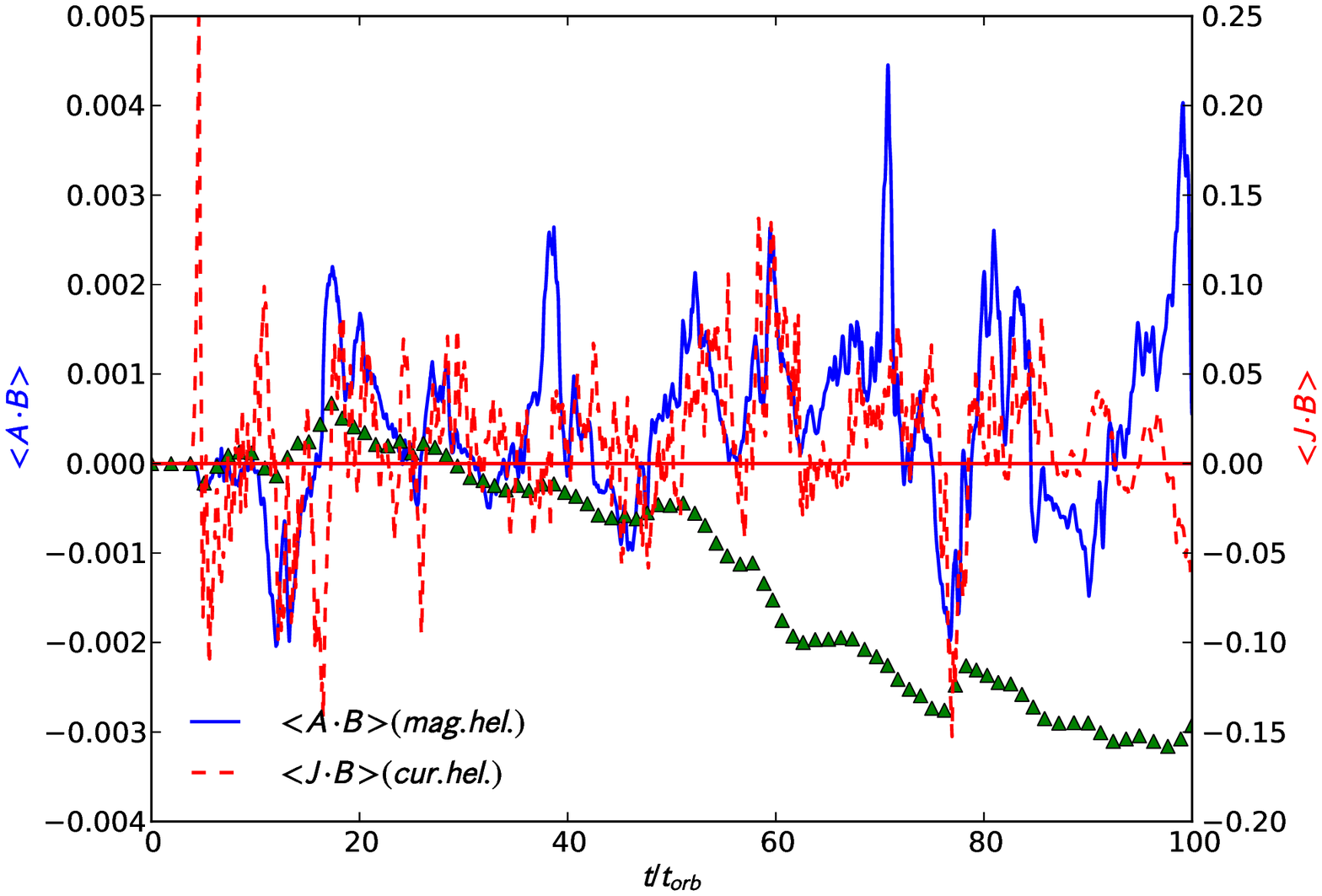}

  \caption{Magnetic helicity $H = \left< \mathbf{A \cdot B} \right>$
    (blue solid line) and current helicity $C = \left< \mathbf{J \cdot
      B} \right>$ (red dashed line), as a function of time for two runs
    with $\Reyn = 3200$ and $\Prandtl = 2$, one with periodic boundary
    conditions (upper panel; ensuring that $H$ is gauge independent
    and thus physically meaningful) and one with VF boundary
    conditions. The green triangles are a simple time integration of
    $-2\eta C$, plotted every 100 timesteps. Horizontal lines mark the
    zero points for each axis. For periodic boundary conditions, the
    integration nearly overlies $H$, demonstrating that magnetic
    helicity is indeed constrained by resistive action on current
    helicity. For VF boundary conditions, the integration (green
    triangles) does not track the helicity at all, because the flux
    terms in Equation~\ref{e:helicity} are not zero in this
    case.  \label{f:abm}}
  
\end{figure}

\begin{figure}
\includegraphics[width=0.95\columnwidth]{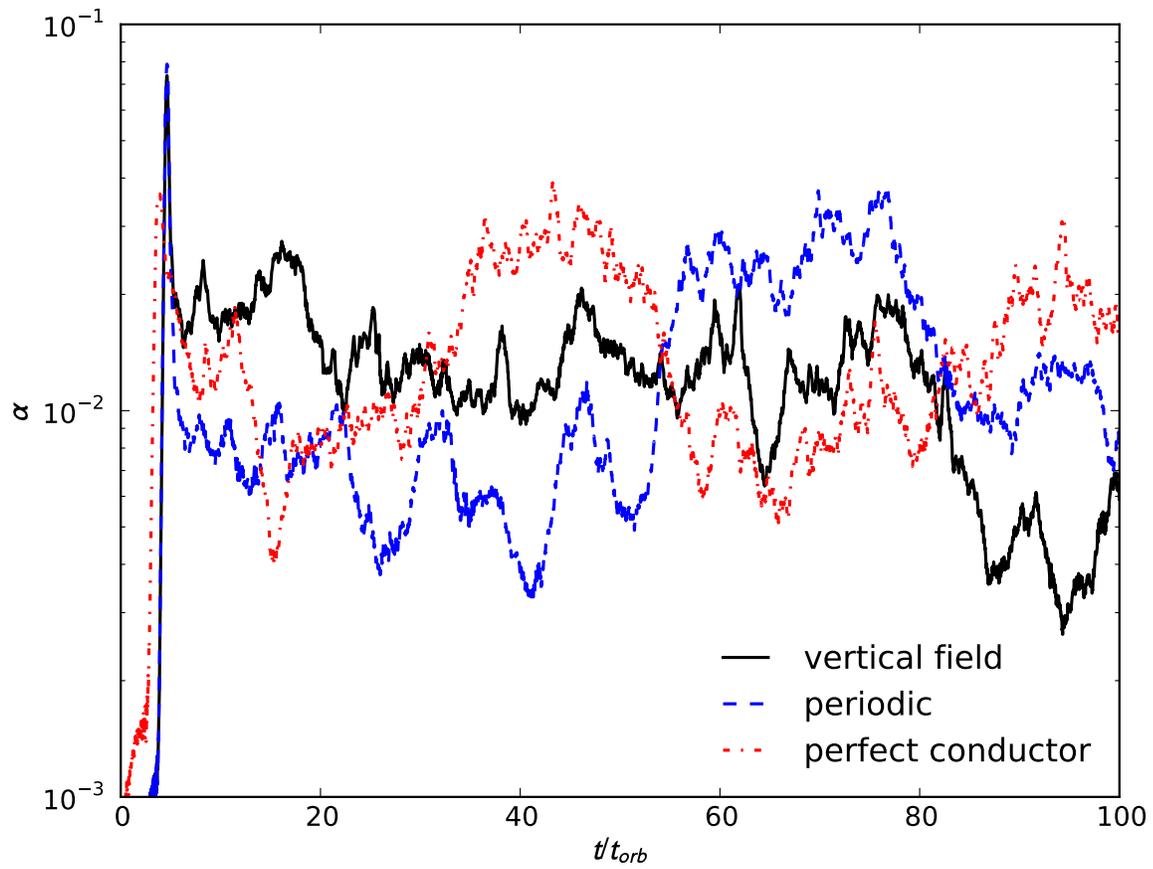}
  \caption{Volume averaged $\left< \alpha \right>$ for a $\Prandtl =
    2$, $\Reyn = 3200$ model with vertical field, periodic, and perfect conductor boundary
    conditions. The transport is comparable regardless of boundary conditions. 
  }
  \label{f:alpha_vs_t_bcs}
\end{figure}

How does transport work in the periodic case, when there is no
boundary flux? The only way for \emph{net} helicity to change in this
case is via resistive effects. Typically, resistively limited systems
lead to lead to catastrophic quenching of the dynamo effect, where
saturated magnetic energy is $\propto \Rmag^{-1/2}$
\citep{1992ApJ...393..165V}. We have shown that this is not the case
in the stratified MRI: boundary conditions do not make significant
differences in the transport (Figure~\ref{f:alpha_vs_t_bcs}) and for
vertical field boundaries, Figure~\ref{f:bym_bxm_vs_t} demonstrates
that the saturated large-scale field strength does not decline
strongly with increasing $\Rmag$. \citet{2001ApJ...550..752V}
suggested a potential solution to this situation: helicity may not
need to be ejected \emph{entirely} across a system boundary (as in,
say, a coronal mass ejection in the Solar dynamo). It could instead be
transported \emph{spectrally}, transferring from small scales to
large. Indeed, \citet{2001ApJ...550..824B} show that exactly this
occurs for a helically driven turbulence simulation with no shear or
rotation--the prototypical $\alpha^2$ dynamo. However, in this case,
while equipartition fields are built even for periodic boundary
conditions, the timescale required to do so is $t_{sat} \propto
\eta^{-1}$: instead of catastrophic quenching for low resistivity,
there is a catastrophic timescale problem instead
\citep{2001ApJ...550..824B}. While our data is not conclusive on the
relationship between $\Rmag$, $\Prandtl$, and the cycle period of
large scale magnetic fields, it certainly does not suggest an inverse
relationship between $t_{sat}$ and $\Rmag$.

The two runs in Figure~\ref{f:abm}, taken together, tell an intriguing
tale: the stratified MRI can transport angular momentum and build
large scale magnetic energy even without a global helicity flux, but
if one is allowed, the system does transport helicity across the
domain boundary. Ultimately, as any boundary condition is in some
sense an approximation of reality, we must understand the details of
the accretion disk dynamo independent of the choice of such
conditions. 

The MRI is difficult to analyze in terms of a simple, two-scale mean
field model: it does not present a single energy injection scale at
which we could expect small scale helicity to be generated
\citep{2010ApJ...713...52D}. This lack of clear scale separation makes
the dynamo coefficients extracted via the test field method rather
noisy and somewhat difficult to interpret, though it has been
performed for the MRI by several groups \citep{2005AN....326..787B,
  2010MNRAS.405...41G}. We forgo the test field method here, since we
should be able to see a crude difference between the current helicity
at the small and large scales if this is in fact what allows dynamo
action, and thus transport. It is also clear comparing
Figures~\ref{f:abm} and \ref{f:bym_bxm_vs_t} that the timescale for
building mean fields for the vertical field boundary condition case is
not related in any obvious way to the flux of magnetic
helicity. Furthermore, we do not see any significant changes in
angular momentum transport with differing boundary conditions that do
and do not allow a helicity flux. We address this in the next section.

\subsection{Helicity power spectra}
We want to understand if scale transfer is responsible for dynamo
action in closed, stratified MRI simulations. To do so, we compute the
spectrum for the relative helicity, $\mathcal{H}$, on spherical shells
in $k$-space, in runs with closed boundary conditions (i.e.,
periodic or perfect conductor). This ensures magnetic helicity is
gauge independent by removing the surface terms in
Equation~\ref{e:helicity}.

\begin{figure}
    \includegraphics[width=0.95\columnwidth]{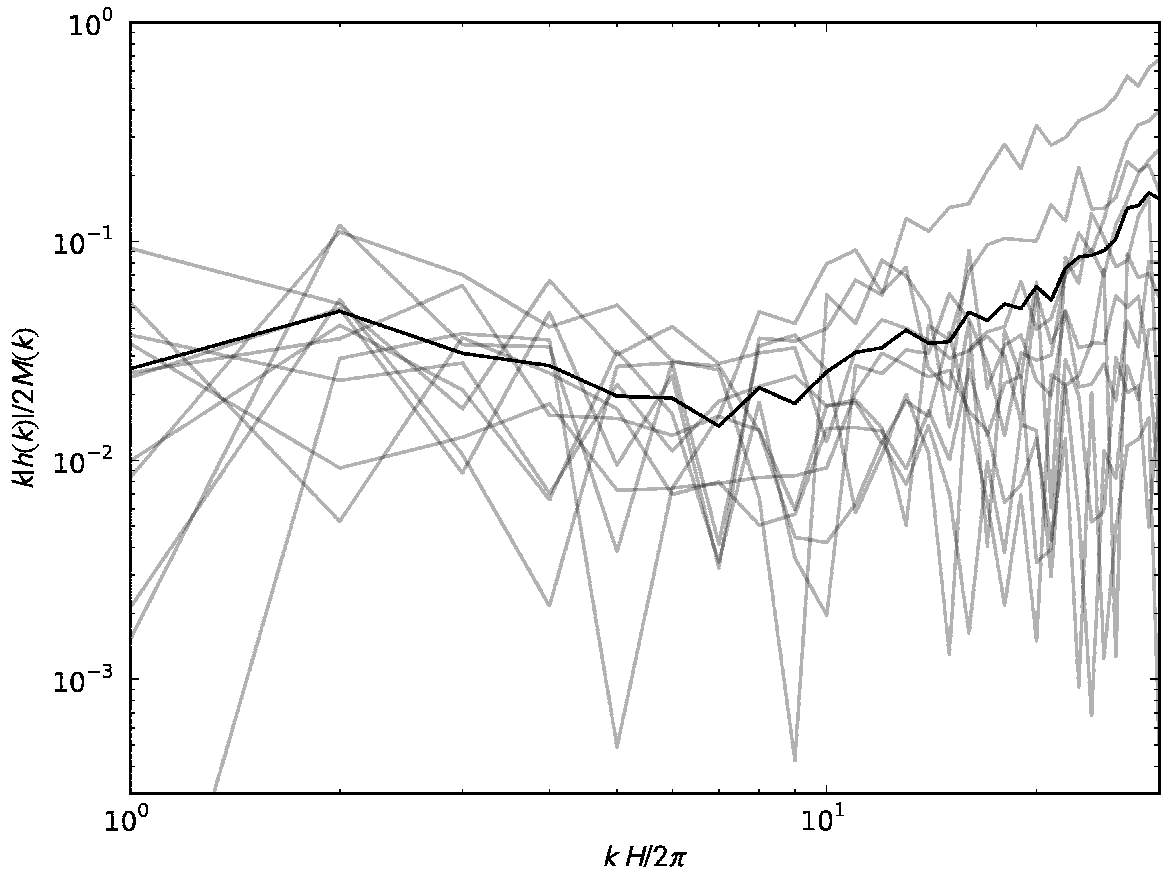}
  \caption{Relative magnetic helicity $\mathcal{H} = k H/ 2 M$ as a function of scale for a run
    with $\Reyn = 3200$, $\Prandtl = 2$ and periodic boundary
    conditions, averaged over a $10 t_{orb}$ period in
    saturation. Light lines show individual timesteps, the dark line
    is the average. The value remains well below unity at all scales,
    showing that the field is not significantly helical at any scale.}
  \label{f:helicity_spectrum}
\end{figure}

In Figure~\ref{f:helicity_spectrum}, we show the helicity spectrum for
$\Reyn = 3200$, $\Prandtl =2$. The field is not strongly helical at
any scale, reaching only $\mathcal{H} \sim 0.2$ at the smallest scales
and remaining much lower on the largest scales. By contrast, the
$\alpha^2$ dynamo driven by a fully helical fluid forcing function has
$\mathcal{H} \sim 1$ at all scales \citep{2001ApJ...550..824B}. In
that case, something similar to the classic inverse cascade of
magnetic energy due to helicity conservation \citep[][and
  \S~\ref{s:helicity}]{1975JFM....68..769F,1976JFM....77..321P}
occurs. In our case, however, we do not see a significant difference
in the properties of the dynamo when the boundary conditions change
between those that do not allow a flux of magnetic helicity and those
that do, despite the fact that our results suggest that the MRI does
in fact eject helicity when given the chance (see
Figure~\ref{f:abm}). This resolves that observation: the MRI generated
field is not strongly helical, even when both kinetic and magnetic
$\alpha$ effects occur. Thus, the constraints placed on the field by
the conservation of magnetic helicity do not dominate its
formation. 

Finally, the fact that the relative helicity is peaked at small scales
suggests that helicity constraints might be more important in the
unstratified MRI, as suggested by \citet{2010arXiv1004.2417K}. In that
case, the generation of field is not affected by dynamo waves and
magnetic buoyancy, and so the helical turbulence might wind the field
into a helically limited state. The small scales in
Figure~\ref{f:helicity_spectrum}, where coherent dynamo waves and
magnetic buoyancy effects are less important, show a rise in helicity,
consistent with this idea.

\subsection{Future Work and Some Speculations}
There remains some incongruity between our results for $\alpha_M(z)$
and those of \citet{2010MNRAS.405...41G}. The only significant
difference between his simulations and ours is his use of outflow
boundary conditions on the fluid (the magnetic field boundary
conditions are identical) and his use of a slightly larger domain. By
running a model with a larger vertical extent, we have shown that
above the $\pm 2 H$ $z$-boundaries of our standard domain, the profile
reverses and matches Gressel's. However, our results show a much
stronger reversal near the midplane than his does, and this appears to
be robust regardless of the domain size. The MRI coupled with outflow
fluid boundary conditions leads to a magnetized wind, and this,
together with the increased efficiency of the Parker instability with
height, likely explains the remaining discrepancy. The role of
\emph{fluid} boundary conditions (which we have not varied here)
should also be examined, in order to better understand how winds
interact with a disk dynamo \citep[e. g.][]{2001ApJ...550..752V}.

The overall pattern of dynamo generated fields in the $z-t$ plane for
the MRI is determined by the boundary conditions
\citep{1995ApJ...446..741B}, and this can be understood in terms of
analytic mean-field theory \citep{2005PhR...417....1B}. Nonetheless,
Figure~\ref{f:alpha_vs_t_bcs} shows that these various dynamo modes do
not affect angular momentum transport. Thus, all of the boundary
conditions studied here for the stratified MRI do an equally good job
of generating magnetic flux for the MRI to continue to grow on. We
have identified the fact that the magnetic fields are not fully
helical at any scale as a likely reason why the boundary conditions do
not determine the total angular momentum transport. Future work should
address this point in more detail. Measuring the helicity spectrum in
unstratified shearing boxes would determine if the dramatic
differences between periodic and VF boundary conditions in that case
are indeed related to a helicity-limited field evolution. Further
theoretical work on the relationship between the mean field dynamo
mechanisms in unstratified domains \citep[likely the incoherent-$\alpha$
  effect of][]{1997ApJ...475..263V} and stratified domains (the $\alpha-\Omega$
effect) and their expected degree of helicity is also required in
order to complete this picture of dynamo-mediated MRI transport in the
zero-net flux case.

We have demonstrated that understanding angular momentum transport via
the MRI is strongly tied to the details of MHD turbulence and
dynamo action. It is worth commenting briefly on some results for
isotropic (non-shearing), non-rotating MHD turbulence, a much better
studied system. Recently, several groups have demonstrated that
non-local interactions in $k$-space are important at large scales in
MHD turbulence, cross-coupling large scale velocity fields with small
scale magnetic fields \citep[e.g.][]{2005PhRvE..72d6302M,
  2009PhRvE..79f6307L, 2010ApJ...725.1786C}. If such an analysis holds
for the stratified MRI dynamo, it could explain the existence of a
critical magnetic Reynolds number $\Rmagc$: if the large scale
velocity fields are coupled to the magnetic dissipation range, a much
more efficient energy sink appears than if they are coupled to a
magnetic inertial range. In the latter case, the large scale
velocities could contribute to small scale helicity production,
continuing to drive large scale dynamo action. This could be verified
by a shell-transfer analysis of the stratified MRI, which we will
pursue in a future publication.

It is possible that our hyperdiffusive terms might be biasing our
results. However, in a series of small-scale dynamo simulations,
groups led by Schekochihin and Brandenburg have found no evidence that
hyperviscosity plays a significant role
\citep{2005ApJ...625L.115S,2007NJPh....9..300S}. There is a
well-understood mechanism by which hyperresistivity causes an increase
in saturated mean field strengths for $\alpha^2$ dynamos
\citep{2002PhRvL..88e5003B}. The main effect of hyperresistivity is
simply to modify the timescales and length scales over which resistive
effects occur, leading to differences from the magnetic helicity
theory developed for Laplacian resistivity
\citep{2001ApJ...550..824B}. Once the hyperresistive scales are
properly accounted for, magnetic helicity conservation indeed
correctly predicts the saturation field strength. Thus, our results are
unlikely to be significantly affected by this mechanism, as we do not
offer a detailed theory for the saturation amplitude for the MRI.
Furthermore, the \citet{2002PhRvL..88e5003B} simulations did not
include a standard, Laplacian resistivity term, while ours do.

More directly, our resolution study also suggests that hyperdiffusion
is not a dominant effect: for integrated quantities, our $64\ 
\mathrm{zones}/H$ and $128\ \mathrm{zones}/H$ runs are converged (see
Figure~\ref{f:alpha_vs_pm} and Table~\ref{t:runs}). Because the
hyperdiffusive coefficients are smaller for larger resolution, this
suggests that the values of the coefficients make little difference.
Nevertheless, we cannot rule out any \emph{systematic} hyperdiffusive
influence on the dynamo behavior without running simulations at high
enough resolution to simultaneously resolve the MRI and dissipate
energy fast enough at the grid scale using only regular Laplacian
diffusion operators at all times and through the entire spatial
domain. While this is practical for unstratified MRI simulations,
because the distribution of $u$ is relatively narrow, in stratified
simulations where the distribution of velocities is much broader, it
is hard to maintain stability without hyperdiffusivity. Direct
numerical simulations of stratified MRI thus require very high
resolution and are therefore extremely expensive and are not practical
for the parameter study considered here.

\section{Conclusions}
\label{s:conclusions}
We have demonstrated that the strength of the angular momentum
transport parameter $\alphaSS$ in stratified, MRI driven turbulence
appears independent of the magnetic Prandtl
number $\Prandtl$ above some critical magnetic Reynolds number $\Rmagc
\sim 3000$. Our models suggest that the value of $\Prandtl$ at which the
flattening of the $\Prandtl - \alphaSS$ relation occurs is a function
of $\Rmag$. \citet{2010arXiv1004.2417K} demonstrated that in the
unstratified case, $\alpha_{SS}$ is independent of $\Prandtl$
if boundary conditions allow for the ejection of magnetic helicity. If
these conclusions are indicative of the asymptotic state in real
disks, then concerns about $\Prandtl$-dependent scaling of the MRI in
real disks, which are certainly stratified, would be entirely
alleviated if the $\Rmag$ is sufficiently high. 

We find that although the stratified MRI does eject magnetic helicity
when boundary conditions allow, the fields it generates are not
significantly helical at any scale or at any time during their
evolution. We suggest that this points toward a theoretical
explanation for why previous stratified MRI simulations with periodic
boundary conditions in the vertical direction sustained dynamo action
while unstratifed simulations at similar $\Prandtl$ did not: while the
\emph{unstratified} MRI dynamo can only grow by ejecting helicity, the
\emph{stratified} MRI dynamo does not wind itself into a fully helical
state and is thus not limited by the transport of magnetic helicity
across boundaries. This is only a preliminary result; future work
should follow in more detail evolution of magnetic helicity in both
unstratified and stratified MRI simulations. The scaling results
presented in this paper appear to be the result of mean field dynamo
action: when $\Rmag > \Rmagc$, the dynamo is ordered and angular
momentum transport does not depend on $\Prandtl$; when $\Rmag <
\Rmagc$, the dynamo is disordered and angular momentum transport is
strongly dependent on $\Prandtl$. This underlines the importance of
understanding the MRI dynamo system in order to understand the details
of angular momentum transport in accretion disks, as pointed out by
\citet{2010AN....331..101B} and \citet{2010MNRAS.405...41G}. In order
to make more concrete predictions about the structure and evolution of
disks, we will need to turn to global simulations. Given that global
simulations inevitably involve much lower resolution than local
simulations, they should involve some kind of sub-grid model in order
to properly capture small-scale dynamics. This work here provides a
preliminary step toward such a model.

Astrophysical disks have tremendous values of $\Reyn$ and $\Rmag$, so even
with a $\Prandtl \sim 10^{-8}$, $\Rmag >> \Rmagc$ if $\Reyn \sim
10^{12}$. In order to make this concrete, we use the
\citet{2008ApJ...674..408B} estimates for the viscosity and
resistivity to estimate the Reynolds number for a typical
protoplanetary disk active region. Using a fiducial midplane density
$\rho \simeq 10^{-10} \mathrm{g\ cm^{-3}}$, scale height $H \simeq
0.05 \mathrm{AU}$, and temperature $T \sim 500 \mathrm{K}$
\citep{2006A&A...445..205I}, we arrive at $\Reyn \simeq 4 \times
10^{16}$. The same estimates give $\Prandtl \simeq 1 \times 10^{-8}$,
easily fulfilling $\Rmag >> \Rmagc$ and so we expect the transport to
be independent of $\Prandtl$ in the \emph{active} regions of such
disks. 

A major caveat to this, of course, is that the ionization state of
such disks is not well established. Our results do not bear on the
question of dead zones, which are regions of high resistivity
resulting from poor ionization rather than low
temperatures. Nevertheless, our point here is to establish the dynamo
state of the active region, \emph{not} to compute detailed predictions
for $\alphaSS$ in such systems. Finally, black hole accretion systems
should also have $\Rmag >> \Rmagc$: the same computation yields
$\Reyn \simeq 6 \times 10^{12}$ and $\Prandtl \simeq 0.1$ at roughly
$100$ Schwarzschild radii from a $10 M_{\odot}$ black hole
\citep{2008ApJ...674..408B}.

\begin{acknowledgments}
We thank Oliver Gressel, Jake Simon, Eliot Quataert, Ian Parrish, Axel
Brandenburg, and Marie Oishi for helpful discussions. We further thank
the anonymous referee for a detailed report that vastly improved our
presentation and discussion. Computations were performed on the Kraken
machine at NICS under grant TG-AST090072, the Big Ben and Pople
systems at Pittsburgh Supercomputing Center under grant TG-MCA99S024,
both supported by NSF, and the NASA Advanced Supercomputing Division's
Pleiades system under grant SMD-19-1846. M\-MML was partly supported by
the NASA Origins of Solar Systems Program under grant NNX07AI74G.
\end{acknowledgments}

\appendix
\section{The Kepler Gauge}
\label{s:gauge}
We here derive the Kepler gauge used in our definition of the magnetic
vector potential. Beginning from the induction equation for the vector
potential,
\begin{equation}
  \frac{\partial \mathbf{A}}{\partial t} = \mathbf{v \times B} - \eta \mu_0 \mathbf{J}
\end{equation}
where $\mathbf{J = \nabla \times B}$, we expand the velocity $\mathbf{v = u + \Ukep}$,
where $\Ukep = q \Omega_0 x \hat{\mathbf{y}}$ is the linearized
Keplerian shear velocity. Making this substitution, we have
\begin{equation}
  \frac{\partial \mathbf{A}}{\partial t} = \mathbf{u} \times \mathbf{B} + [\Ukep \times \mathbf{B}] - \eta
  \mu_0 \mathbf{J}
\end{equation}
Expanding the second term on the right hand side,
\begin{equation}
  \Ukep \times \mathbf{B} = \Ukep \times (\mathbf{\nabla \times A}) = \mathbf{\nabla(\Ukep \cdot A)} - \mathbf{A \times (\nabla \times \Ukep)} - \mathbf{\Ukep \cdot \nabla A} - \mathbf{A \cdot \nabla \Ukep}.
\end{equation}
The first term on the right hand side is the gradient of a scalar function $\Phikep = \mathbf{\Ukep \cdot A}$, which we term the Kepler gauge. The remainder of the terms represent shear and advection, and correspond to the second terms on either side of Equation~\ref{e:induction}. 
We refer the reader to \citet{2011ApJ...727...11H} for a generalization of this gauge to any advective velocity. 


\begin{deluxetable}{lllllllllll}
  \input{table.tex}
  \tablecomments{
    Basic run parameters. $\mathrm{Re'} = u_{rms}/\nu k_1$ and $\mathrm{Rm'} = u_{rms}/\eta k_1$, where $k_1 = 2\pi/4H$ is is the smallest integer wavenumber in the box, measure the ratio of turbulent advection to viscous and magnetic diffusion, respectively. $\eta_3$, $\nu_3$, and $D_3$ are the hyperresistive, hyperviscous, and hyperdiffusive coefficients. 
  }
  \label{t:runs}
\end{deluxetable}

\bibliographystyle{apj}
\bibliography{bibliography}
\end{document}

%% file: table.tex
\tablehead{
\colhead{Run} & \colhead{$\mathrm{Re}$} & \colhead{$\mathrm{Rm}$} & \colhead{$\mathrm{Pm}$} & \colhead{$L_z$ ($H$)}  & \colhead{$\nu_3$, $\eta_3$, $D_3$} & \colhead{zones/H} & \colhead{$\mathrm{Re'}$} & \colhead{$\mathrm{Rm'}$} & \colhead{$u_{rms}$} & \colhead{$B^2/2$}
}
\startdata
A & 12800 & 6400 & 0.5 & 4 & $3.0(-13)$ & 128 & 226.58 & 113.29 &   0.17 &   0.03\\
B & 12800 & 6400 & 0.5 & 4 & $9.6(-12)$ & 64 & 239.96 & 119.98 &   0.18 &   0.03\\
C & 12800 & 12800 & 1 & 4 & $9.6(-12)$ & 64 & 255.15 & 255.15 &   0.19 &   0.04\\
D & 12800 & 25600 & 2 & 4 & $9.6(-12)$ & 64 & 357.16 & 714.32 &   0.26 &   0.05\\
E & 12800 & 51200 & 4 & 4 & $9.6(-12)$ & 64 & 290.71 & 1162.86 &   0.21 &   0.06\\
F & 9600 & 19200 & 2 & 4 & $3.0(-13)$ & 128 & 207.70 & 415.40 &   0.20 &   0.05\\
G & 9600 & 9600 & 1 & 4 & $9.6(-12)$ & 64 & 233.36 & 233.36 &   0.23 &   0.07\\
H & 9600 & 4800 & 0.5 & 4 & $9.6(-12)$ & 64 & 156.13 &  78.06 &   0.15 &   0.02\\
I & 9600 & 2400 & 0.25 & 4 & $9.6(-12)$ & 64 &  99.29 &  24.82 &   0.10 &   0.01\\
J & 6400 & 12800 & 2 & 4 & $9.6(-12)$ & 64 & 132.10 & 264.18 &   0.19 &   0.05\\
K & 6400 & 6400 & 1 & 4 & $9.6(-12)$ & 64 & 103.18 & 103.18 &   0.15 &   0.02\\
L & 6400 & 25600 & 4 & 4 & $9.6(-12)$ & 64 & 180.42 & 721.68 &   0.27 &   0.13\\
Ll & 6400 & 25600 & 4 & 6 & $9.6(-12)$ & 64 & 193.20 & 772.81 &   0.28 &   0.06\\
M & 6400 & 25600 & 4 & 4 & $3.0(-13)$ & 128 & 179.70 & 718.79 &   0.26 &   0.10\\
N & 6400 & 1600 & 0.25 & 4 & $9.6(-12)$ & 64 &  55.96 &  13.99 &   0.08 &   0.01\\
O & 3200 & 12800 & 4 & 4 & $9.6(-12)$ & 64 &  74.08 & 296.31 &   0.22 &   0.06\\
P & 3200 & 6400 & 2 & 4 & $9.6(-12)$ & 64 &  68.31 & 136.62 &   0.20 &   0.05\\
Q & 3200 & 3200 & 1 & 4 & $9.6(-12)$ & 64 &  41.77 &  41.77 &   0.12 &   0.02\\
R & 3200 & 1600 & 0.5 & 4 & $9.6(-12)$ & 64 &  11.24 &   5.62 &   0.03 &   0.01\\
S & 1600 & 6400 & 4 & 4 & $9.6(-12)$ & 64 &  38.27 & 153.07 &   0.23 &   0.08\\
T & 1600 & 3200 & 2 & 4 & $9.6(-12)$ & 64 &  18.22 &  36.43 &   0.11 &   0.02\\
U & 1600 & 1600 & 1 & 4 & $9.6(-12)$ & 64 &  10.96 &  10.96 &   0.06 &   0.02\\
V & 800 & 3200 & 4 & 4 & $9.6(-12)$ & 64 &  16.56 &  66.24 &   0.20 &   0.09\\
W & 800 & 1600 & 2 & 4 & $9.6(-12)$ & 64 &   0.11 &   0.23 &   0.00 &   0.00\\
\enddata

%% file: ms.bbl
\begin{thebibliography}{47}
\expandafter\ifx\csname natexlab\endcsname\relax\def\natexlab#1{#1}\fi

\bibitem[{{Balbus} \& {Hawley}(1992)}]{1992ApJ...400..610B}
{Balbus}, S.~A., \& {Hawley}, J.~F. 1992, \apj, 400, 610

\bibitem[{{Balbus} \& {Hawley}(1998)}]{1998RvMP...70....1B}
---. 1998, Reviews of Modern Physics, 70, 1

\bibitem[{{Balbus} \& {Henri}(2008)}]{2008ApJ...674..408B}
{Balbus}, S.~A., \& {Henri}, P. 2008, \apj, 674, 408

\bibitem[{{Blackman}(2010)}]{2010AN....331..101B}
{Blackman}, E.~G. 2010, Astronomische Nachrichten, 331, 101

\bibitem[{{Brandenburg}(2001)}]{2001ApJ...550..824B}
{Brandenburg}, A. 2001, \apj, 550, 824

\bibitem[{{Brandenburg}(2005)}]{2005AN....326..787B}
---. 2005, Astronomische Nachrichten, 326, 787

\bibitem[{{Brandenburg} \& {Dobler}(2002)}]{2002CoPhC.147..471B}
{Brandenburg}, A., \& {Dobler}, W. 2002, Comp. Phys. Comm., 147, 471

\bibitem[{{Brandenburg} {et~al.}(1995){Brandenburg}, {Nordlund}, {Stein}, \&
  {Torkelsson}}]{1995ApJ...446..741B}
{Brandenburg}, A., {Nordlund}, A., {Stein}, R.~F., \& {Torkelsson}, U. 1995,
  \apj, 446, 741

\bibitem[{{Brandenburg} \& {Sarson}(2002)}]{2002PhRvL..88e5003B}
{Brandenburg}, A., \& {Sarson}, G.~R. 2002, Physical Review Letters, 88, 055003

\bibitem[{{Brandenburg} \& {Subramanian}(2005)}]{2005PhR...417....1B}
{Brandenburg}, A., \& {Subramanian}, K. 2005, \physrep, 417, 1

\bibitem[{{Cho}(2010)}]{2010ApJ...725.1786C}
{Cho}, J. 2010, \apj, 725, 1786

\bibitem[{{Davis} {et~al.}(2010){Davis}, {Stone}, \&
  {Pessah}}]{2010ApJ...713...52D}
{Davis}, S.~W., {Stone}, J.~M., \& {Pessah}, M.~E. 2010, \apj, 713, 52

\bibitem[{{Fleming} {et~al.}(2000){Fleming}, {Stone}, \&
  {Hawley}}]{2000ApJ...530..464F}
{Fleming}, T.~P., {Stone}, J.~M., \& {Hawley}, J.~F. 2000, \apj, 530, 464

\bibitem[{{Frisch} {et~al.}(1975){Frisch}, {Pouquet}, {Leorat}, \&
  {Mazure}}]{1975JFM....68..769F}
{Frisch}, U., {Pouquet}, A., {Leorat}, J., \& {Mazure}, A. 1975, Journal of
  Fluid Mechanics, 68, 769

\bibitem[{{Fromang} \& {Papaloizou}(2007)}]{2007A&A...476.1113F}
{Fromang}, S., \& {Papaloizou}, J. 2007, \aap, 476, 1113

\bibitem[{{Fromang} {et~al.}(2007){Fromang}, {Papaloizou}, {Lesur}, \&
  {Heinemann}}]{2007A&A...476.1123F}
{Fromang}, S., {Papaloizou}, J., {Lesur}, G., \& {Heinemann}, T. 2007, \aap,
  476, 1123

\bibitem[{{Gressel}(2010)}]{2010MNRAS.405...41G}
{Gressel}, O. 2010, \mnras, 405, 41

\bibitem[{{Hawley} {et~al.}(1995){Hawley}, {Gammie}, \&
  {Balbus}}]{1995ApJ...440..742H}
{Hawley}, J.~F., {Gammie}, C.~F., \& {Balbus}, S.~A. 1995, \apj, 440, 742

\bibitem[{{Hawley} {et~al.}(1996){Hawley}, {Gammie}, \&
  {Balbus}}]{1996ApJ...464..690H}
---. 1996, \apj, 464, 690

\bibitem[{{Hubbard} \& {Brandenburg}(2011)}]{2011ApJ...727...11H}
{Hubbard}, A., \& {Brandenburg}, A. 2011, \apj, 727, 11

\bibitem[{{Hurlburt} \& {Toomre}(1988)}]{1988ApJ...327..920H}
{Hurlburt}, N.~E., \& {Toomre}, J. 1988, \apj, 327, 920

\bibitem[{{Ilgner} \& {Nelson}(2006)}]{2006A&A...445..205I}
{Ilgner}, M., \& {Nelson}, R.~P. 2006, \aap, 445, 205

\bibitem[{{Johansen} \& {Klahr}(2005)}]{2005ApJ...634.1353J}
{Johansen}, A., \& {Klahr}, H. 2005, \apj, 634, 1353

\bibitem[{{Johansen} {et~al.}(2009){Johansen}, {Youdin}, \&
  {Klahr}}]{2009ApJ...697.1269J}
{Johansen}, A., {Youdin}, A., \& {Klahr}, H. 2009, \apj, 697, 1269

\bibitem[{{K{\"a}pyl{\"a}} \& {Korpi}(2010)}]{2010arXiv1004.2417K}
{K{\"a}pyl{\"a}}, P.~J., \& {Korpi}, M.~J. 2010, ArXiv e-prints

\bibitem[{{Kitchatinov} \& {R{\"u}diger}(2010)}]{2010A&A...513L...1K}
{Kitchatinov}, L.~L., \& {R{\"u}diger}, G. 2010, \aap, 513, L1

\bibitem[{{Lessinnes} {et~al.}(2009){Lessinnes}, {Carati}, \&
  {Verma}}]{2009PhRvE..79f6307L}
{Lessinnes}, T., {Carati}, D., \& {Verma}, M.~K. 2009, \pre, 79, 066307

\bibitem[{{Lesur} \& {Longaretti}(2007)}]{2007MNRAS.378.1471L}
{Lesur}, G., \& {Longaretti}, P.-Y. 2007, \mnras, 378, 1471

\bibitem[{{Lesur} \& {Ogilvie}(2008)}]{2008A&A...488..451L}
{Lesur}, G., \& {Ogilvie}, G.~I. 2008, \aap, 488, 451

\bibitem[{{Longaretti} \& {Lesur}(2010)}]{2010A&A...516A..51L}
{Longaretti}, P.-Y., \& {Lesur}, G. 2010, \aap, 516, A51

\bibitem[{{Mininni} {et~al.}(2005){Mininni}, {Alexakis}, \&
  {Pouquet}}]{2005PhRvE..72d6302M}
{Mininni}, P., {Alexakis}, A., \& {Pouquet}, A. 2005, \pre, 72, 046302

\bibitem[{{Oishi} {et~al.}(2007){Oishi}, {Mac Low}, \&
  {Menou}}]{2007ApJ...670..805O}
{Oishi}, J.~S., {Mac Low}, M., \& {Menou}, K. 2007, \apj, 670, 805

\bibitem[{{Pessah}(2010)}]{2010ApJ...716.1012P}
{Pessah}, M.~E. 2010, \apj, 716, 1012

\bibitem[{{Pessah} {et~al.}(2007){Pessah}, {Chan}, \&
  {Psaltis}}]{2007ApJ...668L..51P}
{Pessah}, M.~E., {Chan}, C.-k., \& {Psaltis}, D. 2007, \apjl, 668, L51

\bibitem[{{Pouquet} {et~al.}(1976){Pouquet}, {Frisch}, \&
  {Leorat}}]{1976JFM....77..321P}
{Pouquet}, A., {Frisch}, U., \& {Leorat}, J. 1976, Journal of Fluid Mechanics,
  77, 321

\bibitem[{{Regev} \& {Umurhan}(2008)}]{2008A&A...481...21R}
{Regev}, O., \& {Umurhan}, O.~M. 2008, \aap, 481, 21

\bibitem[{{Sano} \& {Stone}(2002)}]{2002ApJ...577..534S}
{Sano}, T., \& {Stone}, J.~M. 2002, \apj, 577, 534

\bibitem[{{Schekochihin} {et~al.}(2005){Schekochihin}, {Haugen}, {Brandenburg},
  {Cowley}, {Maron}, \& {McWilliams}}]{2005ApJ...625L.115S}
{Schekochihin}, A.~A., {Haugen}, N.~E.~L., {Brandenburg}, A., {Cowley}, S.~C.,
  {Maron}, J.~L., \& {McWilliams}, J.~C. 2005, \apjl, 625, L115

\bibitem[{{Schekochihin} {et~al.}(2007){Schekochihin}, {Iskakov}, {Cowley},
  {McWilliams}, {Proctor}, \& {Yousef}}]{2007NJPh....9..300S}
{Schekochihin}, A.~A., {Iskakov}, A.~B., {Cowley}, S.~C., {McWilliams}, J.~C.,
  {Proctor}, M.~R.~E., \& {Yousef}, T.~A. 2007, New Journal of Physics, 9, 300

\bibitem[{{Shakura} \& {Sunyaev}(1973)}]{1973A&A....24..337S}
{Shakura}, N.~I., \& {Sunyaev}, R.~A. 1973, \aap, 24, 337

\bibitem[{{Shi} {et~al.}(2010){Shi}, {Krolik}, \&
  {Hirose}}]{2010ApJ...708.1716S}
{Shi}, J., {Krolik}, J.~H., \& {Hirose}, S. 2010, \apj, 708, 1716

\bibitem[{{Simon} {et~al.}(2011){Simon}, {Hawley}, \&
  {Beckwith}}]{2011ApJ...730...94S}
{Simon}, J.~B., {Hawley}, J.~F., \& {Beckwith}, K. 2011, \apj, 730, 94

\bibitem[{{Vainshtein} \& {Cattaneo}(1992)}]{1992ApJ...393..165V}
{Vainshtein}, S.~I., \& {Cattaneo}, F. 1992, \apj, 393, 165

\bibitem[{{Vishniac}(2009)}]{2009ApJ...696.1021V}
{Vishniac}, E.~T. 2009, \apj, 696, 1021

\bibitem[{{Vishniac} \& {Brandenburg}(1997)}]{1997ApJ...475..263V}
{Vishniac}, E.~T., \& {Brandenburg}, A. 1997, \apj, 475, 263

\bibitem[{{Vishniac} \& {Cho}(2001)}]{2001ApJ...550..752V}
{Vishniac}, E.~T., \& {Cho}, J. 2001, \apj, 550, 752

\bibitem[{{Ziegler} \& {R{\"u}diger}(2001)}]{2001A&A...378..668Z}
{Ziegler}, U., \& {R{\"u}diger}, G. 2001, \aap, 378, 668

\end{thebibliography}


\begin{thebibliography}{35}
\expandafter\ifx\csname natexlab\endcsname\relax\def\natexlab#1{#1}\fi

\bibitem[{{Balbus} \& {Hawley}(1998)}]{1998RvMP...70....1B}
{Balbus}, S.~A., \& {Hawley}, J.~F. 1998, Reviews of Modern Physics, 70, 1

\bibitem[{{Balbus} \& {Henri}(2008)}]{2008ApJ...674..408B}
{Balbus}, S.~A., \& {Henri}, P. 2008, \apj, 674, 408

\bibitem[{{Brandenburg}(2001)}]{2001ApJ...550..824B}
{Brandenburg}, A. 2001, \apj, 550, 824

\bibitem[{{Brandenburg} \& {Dobler}(2002)}]{2002CoPhC.147..471B}
{Brandenburg}, A., \& {Dobler}, W. 2002, Comp. Phys. Comm., 147, 471

\bibitem[{{Brandenburg} {et~al.}(1995){Brandenburg}, {Nordlund}, {Stein}, \&
  {Torkelsson}}]{1995ApJ...446..741B}
{Brandenburg}, A., {Nordlund}, A., {Stein}, R.~F., \& {Torkelsson}, U. 1995,
  \apj, 446, 741

\bibitem[{{Brandenburg} \& {Subramanian}(2005)}]{2005PhR...417....1B}
{Brandenburg}, A., \& {Subramanian}, K. 2005, \physrep, 417, 1

\bibitem[{{Cho}(2010)}]{2010ApJ...725.1786C}
{Cho}, J. 2010, \apj, 725, 1786

\bibitem[{{Davis} {et~al.}(2010){Davis}, {Stone}, \&
  {Pessah}}]{2010ApJ...713...52D}
{Davis}, S.~W., {Stone}, J.~M., \& {Pessah}, M.~E. 2010, \apj, 713, 52

\bibitem[{{Frisch} {et~al.}(1975){Frisch}, {Pouquet}, {Leorat}, \&
  {Mazure}}]{1975JFM....68..769F}
{Frisch}, U., {Pouquet}, A., {Leorat}, J., \& {Mazure}, A. 1975, Journal of
  Fluid Mechanics, 68, 769

\bibitem[{{Fromang} \& {Papaloizou}(2007)}]{2007A&A...476.1113F}
{Fromang}, S., \& {Papaloizou}, J. 2007, \aap, 476, 1113

\bibitem[{{Fromang} {et~al.}(2007){Fromang}, {Papaloizou}, {Lesur}, \&
  {Heinemann}}]{2007A&A...476.1123F}
{Fromang}, S., {Papaloizou}, J., {Lesur}, G., \& {Heinemann}, T. 2007, \aap,
  476, 1123

\bibitem[{{Gressel}(2010)}]{2010MNRAS.405...41G}
{Gressel}, O. 2010, \mnras, 405, 41

\bibitem[{{Hawley} {et~al.}(1995){Hawley}, {Gammie}, \&
  {Balbus}}]{1995ApJ...440..742H}
{Hawley}, J.~F., {Gammie}, C.~F., \& {Balbus}, S.~A. 1995, \apj, 440, 742

\bibitem[{{Hawley} {et~al.}(1996){Hawley}, {Gammie}, \&
  {Balbus}}]{1996ApJ...464..690H}
---. 1996, \apj, 464, 690

\bibitem[{{Hubbard} \& {Brandenburg}(2010)}]{2010GApFD.104..577H}
{Hubbard}, A., \& {Brandenburg}, A. 2010, Geophysical and Astrophysical Fluid
  Dynamics, 104, 577

\bibitem[{{Hubbard} \& {Brandenburg}(2011)}]{2011ApJ...727...11H}
---. 2011, \apj, 727, 11

\bibitem[{{Ilgner} \& {Nelson}(2006)}]{2006A&A...445..205I}
{Ilgner}, M., \& {Nelson}, R.~P. 2006, \aap, 445, 205

\bibitem[{{Johansen} {et~al.}(2009){Johansen}, {Youdin}, \&
  {Klahr}}]{2009ApJ...697.1269J}
{Johansen}, A., {Youdin}, A., \& {Klahr}, H. 2009, \apj, 697, 1269

\bibitem[{{K{\"a}pyl{\"a}} \& {Korpi}(2010)}]{2010arXiv1004.2417K}
{K{\"a}pyl{\"a}}, P.~J., \& {Korpi}, M.~J. 2010, ArXiv e-prints

\bibitem[{{Kitchatinov} \& {R{\"u}diger}(2010)}]{2010A&A...513L...1K}
{Kitchatinov}, L.~L., \& {R{\"u}diger}, G. 2010, \aap, 513, L1+

\bibitem[{{Lessinnes} {et~al.}(2009){Lessinnes}, {Carati}, \&
  {Verma}}]{2009PhRvE..79f6307L}
{Lessinnes}, T., {Carati}, D., \& {Verma}, M.~K. 2009, \pre, 79, 066307

\bibitem[{{Lesur} \& {Longaretti}(2007)}]{2007MNRAS.378.1471L}
{Lesur}, G., \& {Longaretti}, P.-Y. 2007, \mnras, 378, 1471

\bibitem[{{Lesur} \& {Ogilvie}(2008)}]{2008A&A...488..451L}
{Lesur}, G., \& {Ogilvie}, G.~I. 2008, \aap, 488, 451

\bibitem[{{Mininni} {et~al.}(2005){Mininni}, {Alexakis}, \&
  {Pouquet}}]{2005PhRvE..72d6302M}
{Mininni}, P., {Alexakis}, A., \& {Pouquet}, A. 2005, \pre, 72, 046302

\bibitem[{{Pessah} {et~al.}(2007){Pessah}, {Chan}, \&
  {Psaltis}}]{2007ApJ...668L..51P}
{Pessah}, M.~E., {Chan}, C., \& {Psaltis}, D. 2007, \apjl, 668, L51

\bibitem[{{Pouquet} {et~al.}(1976){Pouquet}, {Frisch}, \&
  {Leorat}}]{1976JFM....77..321P}
{Pouquet}, A., {Frisch}, U., \& {Leorat}, J. 1976, Journal of Fluid Mechanics,
  77, 321

\bibitem[{{Regev} \& {Umurhan}(2008)}]{2008A&A...481...21R}
{Regev}, O., \& {Umurhan}, O.~M. 2008, \aap, 481, 21

\bibitem[{{Schekochihin} {et~al.}(2005){Schekochihin}, {Haugen}, {Brandenburg},
  {Cowley}, {Maron}, \& {McWilliams}}]{2005ApJ...625L.115S}
{Schekochihin}, A.~A., {Haugen}, N.~E.~L., {Brandenburg}, A., {Cowley}, S.~C.,
  {Maron}, J.~L., \& {McWilliams}, J.~C. 2005, \apjl, 625, L115

\bibitem[{{Schekochihin} {et~al.}(2007){Schekochihin}, {Iskakov}, {Cowley},
  {McWilliams}, {Proctor}, \& {Yousef}}]{2007NJPh....9..300S}
{Schekochihin}, A.~A., {Iskakov}, A.~B., {Cowley}, S.~C., {McWilliams}, J.~C.,
  {Proctor}, M.~R.~E., \& {Yousef}, T.~A. 2007, New Journal of Physics, 9, 300

\bibitem[{{Shakura} \& {Sunyaev}(1973)}]{1973A&A....24..337S}
{Shakura}, N.~I., \& {Sunyaev}, R.~A. 1973, \aap, 24, 337

\bibitem[{{Shi} {et~al.}(2010){Shi}, {Krolik}, \&
  {Hirose}}]{2010ApJ...708.1716S}
{Shi}, J., {Krolik}, J.~H., \& {Hirose}, S. 2010, \apj, 708, 1716

\bibitem[{{Simon} {et~al.}(2010){Simon}, {Hawley}, \&
  {Beckwith}}]{2010arXiv1010.0005S}
{Simon}, J.~B., {Hawley}, J.~F., \& {Beckwith}, K. 2010, \apj submitted

\bibitem[{{Vainshtein} \& {Cattaneo}(1992)}]{1992ApJ...393..165V}
{Vainshtein}, S.~I., \& {Cattaneo}, F. 1992, \apj, 393, 165

\bibitem[{{Vishniac}(2009)}]{2009ApJ...696.1021V}
{Vishniac}, E.~T. 2009, \apj, 696, 1021

\bibitem[{{Vishniac} \& {Cho}(2001)}]{2001ApJ...550..752V}
{Vishniac}, E.~T., \& {Cho}, J. 2001, \apj, 550, 752

\end{thebibliography}
